\documentclass[aps,prd,reprint,nofootinbib, preprintnumbers, superscriptaddress]{revtex4-1}

\usepackage{bm}
\usepackage{graphicx} %
\usepackage{amsmath} %
\usepackage{amssymb} %
\usepackage{url}
\usepackage{subfigure} %
\usepackage{float} %
\usepackage{upgreek} %
\usepackage{multirow} %
\usepackage{color} %
\usepackage{lineno} %
\usepackage{hyperref} %
\usepackage{booktabs}
\usepackage{placeins}
\usepackage{ulem}

\usepackage[usenames,dvipsnames]{xcolor}
\hypersetup{colorlinks=true, urlcolor=black, linkcolor=blue, citecolor=red} %

\usepackage{amsmath}
\usepackage{tcolorbox}
\allowdisplaybreaks[4]

\begin{document}

\title{One-loop tensor power spectrum from an excited scalar field during inflation}

\author{Atsuhisa Ota}
\email{iasota@ust.hk}
\affiliation {HKUST Jockey Club Institute for Advanced Study, The Hong Kong University of Science and Technology, Clear Water Bay, Hong Kong, P.R.China}

\author{Misao Sasaki}
\affiliation{Kavli Institute for the Physics and Mathematics of the Universe (WPI), University of Tokyo, Chiba, 277-8583, Japan}
\affiliation{Center for Gravitational Physics and Quantum Information, Yukawa Institute for Theoretical Physics, Kyoto University, Kyoto, 606-8502, Japan}
\affiliation{Leung Center for Cosmology and Particle Astrophysics, National Taiwan University, Taipei 10617}
\author{Yi Wang}
\affiliation{Department of Physics, The Hong Kong University of Science and Technology, Clear Water Bay, Hong Kong, P.R.China }
\affiliation {HKUST Jockey Club Institute for Advanced Study, The Hong Kong University of Science and Technology, Clear Water Bay, Hong Kong, P.R.China}

\date{\today}

\begin{abstract}
We present a consistent one-loop calculation for the inflationary tensor power spectrum in the presence of an excited spectator scalar field using the in-in formalism. We find that the super-horizon primordial power spectrum of the tensor mode can be scale-invariantly enhanced or reduced by the loop effects of a subhorizon scalar field.
Our calculation also includes the scalar-induced gravitational wave spectrum classically computed in the previous literature, which is significant only near the scales where the scalar field is amplified.
The super-horizon enhancement is a higher-order effect of the interaction Hamiltonian, which can be understood as a Bogoliubov transformation introduced by nonlinear interactions. 
On the other hand, the scale-invariant reduction of the tensor power spectrum may occur due to the fourth-order scalar-scalar-tensor-tensor coupling.
This phenomenon can  be understood as the evolution of an anisotropic Bianchi type-I background in the separate universe approach.
Our result suggests that large-scale measurements may indirectly test the dramatic effects of small-scale cosmological perturbations through loop corrections. This possibility opens a new ground in probing the small-scale physics of the primordial Universe through gravitational wave detectors of cosmological scales.
%
%
%
%
%
%
%
%
\end{abstract}

\maketitle

\section{Introduction}

The one-loop corrections to the inflationary power spectra, and their UV and IR divergences are important components in cosmological perturbation theory of the primordial universe~\cite{Weinberg:2005vy,Urakawa:2008rb,Senatore:2009cf,delRio:2018vrj,Tan:2019czo,Comelli:2022ikb,Comelli:2022ikb,Dimastrogiovanni:2022afr}.
Most studies focus on tiny loop corrections suppressed by the Hubble-to-Planck-mass ratio  $H^2/M^2_{\rm pl}\sim 10^{-10}\epsilon$ with the first slow roll parameter $\epsilon\equiv -\dot H / H$.
Thus it seems the motivation of such loop analyses is to understand what our theories entail, rather than to obtain experimentally verifiable results.

Recent progress in cosmological observations may shed new light on the situation.
The Laser Interferometer Gravitational-Wave Observatory (LIGO) and Virgo event in 2015 reported an unexpectedly massive black hole~\cite{LIGOScientific:2016aoc}, which rejuvenated the idea of primordial black holes~(PBHs)~\cite{Sasaki:2016jop, Clesse:2016vqa, Bird:2016dcv}.
PBHs are formed from collapse of Hubble horizon size regions caused by large curvature perturbations~\cite{Hawking:1971ei,Carr:1974nx,Carr:1975qj}; therefore the existence of PBHs suggests scale-dependence of scalar fluctuations and thus drastic enhancement of quantum fields at some point during inflation~\cite{Alabidi:2012ex,Drees:2011hb,Drees:2011yz,Garcia-Bellido:2017mdw,Ivanov:1994pa,Ezquiaga:2017fvi,Kannike:2017bxn,Germani:2017bcs,Motohashi:2017kbs,Yokoyama:1998pt,Saito:2008em,Zhou:2020kkf,Chen:2019zza,Pi:2021dft}.
UV or IR divergences are no longer expected in loop integrals if one focuses on contributions from excited states of a quantum field.
Instead, an excited quantum field may introduce sizable and observable loop corrections.
Thus, loop corrections motivated by PBHs may lead to interesting phenomenology.
The purpose of this paper is to present a consistent one-loop calculation for such an excited state during inflation. 

The inflationary loops arise from nonlinear interactions of cosmological perturbations.
In general, interactions of scalars are model-dependent and thus complicated. In contrast, the coupling to the tensor fluctuations is limited to the kinetic term regardless of the interaction among scalars for minimally coupled scalar fields in general relativity.
Hence, the possible loop corrections to the tensor power spectrum are easier to deal with than those in the scalar case.
Moreover, ongoing and future gravitational wave (GW) measurements are attracting growing attention from observational perspectives. 
Therefore, we focus on the one-loop corrections to the tensor power spectrum~(see e.g., Refs.~\cite{Kristiano:2022maq,Inomata:2022yte} for recent works on scalar inflationary loops from excited scalar fields.).
The appropriate formalism for the present purpose is a full quantum approach, known as the ``in-in formalism,'' as we consider the quantum regime of inflationary fluctuations.

One often considers a secondary GW production from cosmological perturbations (see, e.g., Ref.~\cite{Domenech:2021ztg} and references therein), where the classical equation of motion for GWs with a source is integrated. The source usually consists of a bilinear form of stochastic variables that represent the scalar-type cosmological perturbation.
The induced tensor fluctuations are intensely investigated as their detection may be a smoking gun for the PBH formation in the very early Universe. 
Conventionally the induced GW production is understood from a viewpoint of classical nonlinear dynamics.
However, the same process may also be regarded as leading order quantum corrections to the tensor modes from a viewpoint of quantum field theory.
From this perspective, we point out that the computation of the induced GWs corresponds to that in the Born approximation of the leading order interaction Hamiltonian.
Thus it is important to understand effects beyond the Born approximation, which include in particular the iterative (or higher-loop) corrections arising from the leading order interaction Hamiltonian as well as those from higher-order interaction Hamiltonians.

Our recent letter~\cite{Ota:2022hvh} reported several surprising results when including the iterative corrections as well as a higher order Hamiltonian in the in-in formalism.
We found that the super-horizon tensor power spectrum may be scale-invariantly enhanced or reduced by the loop effect, which can be understood as a Bogoliubov transformation introduced by the nonlinear interactions. 
This paper provides detailed derivations of the result and an analysis of a new model not discussed in Ref.~\cite{Ota:2022hvh}, including new perspectives such as the separate universe approach.

\medskip
We organize this paper as follows.
In section~\ref{sec2:w}, we review the previous approach to GW productions with stochastic variables and point out two important missing components in it.
In Section~\ref{ininrev}, we provide a brief but self-contained review of the in-in formalism for the one-loop calculation of the tensor power spectrum.
Section~\ref{inf-loop} describes the cosmological setup in this paper.
We apply the in-in formulas to inflationary calculations.
Then in section~\ref{loopfromex}, we derive the loop spectra from an excited scalar field.
In section~\ref{BTsec}, we discuss physical implications of the results 
from a perspective of the Bogoliubov transformation.
In Section~\ref{bianchi1}, we discuss the separate universe approach to obtain a more intuitive understanding of our result for the scale invariant Born approximation contribution.
Conclusions are given in Section~\ref{conc}, and we supplemented some mathematical details in the appendices.

\section{What is new?}\label{sec2:w}

In cosmology, one often computes gravitational waves in classical field theory.
The classical equation of motion for the cosmological tensor perturbation $h_{ij}$ with source is obtained by perturbatively expanding the Einstein equation as
\begin{align}
	\frac{\partial^2 h_{ij}}{\partial \tau^2} + 2\mathcal H \frac{\partial h_{ij}}{\partial \tau} -\nabla^2h_{ij} =\Pi_{ij}{}^{kl}S_{kl}\,,\label{eomh}
\end{align}
where $\tau$ is the conformal time, and $\mathcal H$ is the conformal Hubble parameter, 
${\mathcal H}=d\ln(a)/d\tau$.
The source $S_{ij}$ is obtained by expanding the Einstein tensor and the energy-momentum tensor, and $\Pi_{ij}{}^{kl}$ is the projection operator onto the transverse-traceless subspace.
Given the Green's function $G$ with an appropriate boundary condition,
Eq.~\eqref{eomh} is formally integrated to give
\begin{align}
 h_{ij}(x) &=h_{{\rm hom},ij}(x) +h_{{\rm ind},ij}(x)\,;
 \nonumber\\
 h_{{\rm ind},ij}(x)&=\int d^4x' G(x;x') \Sigma_{ij}(x'),\label{clsol}
\end{align}
where $x$ and $x'$ are 4-dimensional coordinates and $\Sigma_{ij} \equiv \Pi_{ij}{}^{kl}S_{kl}$,
where $h_{{\rm hom},ij}$ is a homogeneous solution. With the retarded boundary condition on $G$, it is the primordial tensor perturbation from inflation in the standard scenario of the early universe.
The inhomogeneous part represents gravitational waves produced, e.g., by the nonlinear evolution of density fluctuations.
The inhomogeneous part $h_{{\rm ind},ij}$ in Eq.~\eqref{clsol} may be regarded as a classical field, but the source configuration is random, as the initial condition of cosmological perturbations may originate from quantum fluctuations during inflation.
Hence $h_{{\rm ind},ij}$ should also be random, whose statistical property is characterized by correlation functions.
In cosmology, we often consider the 2-point correlation function,
\begin{align}
	&\left\langle h_{ij}(x)h^{ij}(y) \right \rangle = 
 \left\langle h_{{\rm hom},ij}(x)h_{\rm hom}^{ij}(y) \right \rangle 
	\notag \\
	&+ \int d^4x'd^4y' G(x;x')G(y;y')\langle \Sigma_{ij}(x')\Sigma^{ij}(y')\rangle, \label{hhinhomo} 
\end{align}
where the first term due to the homogeneous part is usually assumed to be negligible.
Previous literature considered various origins of $\Sigma_{ij}$ and the associated secondary GWs.
This is a standard prescription to compute GWs from stochastic sources.

We mention that there are two additional important contributions 
that must be included in the above computation~(see also Ref.~\cite{Chen:2022dah}.).
First, the cross term of homogeneous and inhomogeneous solutions is missing in Eq.~\eqref{hhinhomo}.
In addition to the auto-correlation of $h_{{\rm hom},ij}$, the following cross term 
may be non-vanishing:
\begin{align}
	\left\langle h_{ij}(x)h^{ij}(y) \right \rangle  
	\supset &  \int d^4y' G(y;y')\langle  h_{{\rm hom},ij}(x)\Sigma^{ij}(y')\rangle.
\end{align}
In fact, it is in general non-vanishing if we expand the source to third order in perturbation.
At leading order, the source is of the second order in perturbation, and so is the induced part of $h_{ij}$. Therefore, the third-order term is sub-leading in $h_{ij}$.
However, the order of the cross-correlation between the first- and third-order terms
is equivalent to that of the auto-correlation of the second-order term.
Thus the cross-correlation may also equally contribute to the correlation function.
In the language of quantum theory, ignoring the cross term implies an incomplete loop expansion.

The second missing contribution is from the iterative solution.
The formal solution \eqref{eomh} is well-defined in the Born approximation.
However, once we take nonlinear terms into account, the source term will contain the tensor perturbation that appears on the left-hand side. 
Perturbatively, this means we have to solve it iteratively. 
In particular, the iterative solution to the next-to-leading order in the source needs to be included for consistency. Namely, like the first issue, the first iterative correction is sub-leading at the field level but may contribute to the correlation function at the same order as the
conventional induced GW part.

To summarize, we need to take into account all possible higher-order interactions and iterative solutions to consistently calculate the nonlinear corrections to cosmological correlation functions. In this paper, we perform a consistent computation up to the one-loop order.
We mention that similar classical stochastic loops are discussed in the nonlinear matter power spectrum of large-scale structure~\cite{Bernardeau:2001qr}.
In that context, all effects are included consistently.

The standard approach based on Eq.~\eqref{eomh} applies to classical stochastic systems.
In cosmology, the method should explain the evolution of random fields after inflation~\cite{Chen:2022dah}.
We are interested in the quantum phase during inflation in this paper, so we consider the in-in formalism instead.
As discussed in Section~\ref{BTsec}, the quantum evolution of the field operator $h_{ij}$ in the interaction picture is expanded into
\begin{align}
	&h_{ij} =h_{{\rm I},ij} +  i \lambda \int^\tau_{\tau_0} d\tau' [H_{{\rm int},I}(\tau'),h_{{\rm I},ij}]
	\notag \\
	&
-\lambda^2 \int^\tau_{\tau_0} d\tau' \int^{\tau'}_{\tau_0} d\tau'' [H_{{\rm int},I}(\tau''),[H_{{\rm int},I}(\tau '),h_{{\rm I},ij}]]
\notag \\
&+\mathcal O(\lambda^3),
\label{hijhei}
\end{align}
where $h_{{\rm I},ij}$ is the interaction picture field, i.e., the linear tensor fluctuations, and $H_{{\rm int}, I}$ is the interaction Hamiltonian in the interaction picture.
$\tau_0$ is the initial time, and the $\tau$ dependence is suppressed for notational simplicity.
Order in $\lambda$ implies the number of vertecies, and the standard induced GWs are the leading order correction in the first line.
The power of $H^2/M_{\rm pl}^2$ counts the number of inflationary loops.
The one-loop correction to the 2-point function contains terms up to $\mathcal O(\lambda^2)$, so there is no prior reason to truncate the expansion at $\mathcal O(\lambda)$ in Eq.~\eqref{hijhei}.

(In)equivalence of the classical stochastic approach and the quantum in-in formalism during inflation is not obvious. A classical stochastic theory can also describe a class of quantum systems, as Nelson discussed in Ref.~\cite{Nelson:1966sp}.
The comparison of the two approaches will be discussed in a separate paper in future work.
In the next section, we start with a lightning review of the in-in formalism.

\section{In-in formalism: a review}\label{ininrev}

We are interested in the time evolution of the vacuum expectation value~(VEV) of a field operator $O$ during inflation.
In the Heisenberg picture, the VEV should be written as  
\begin{align}
	\langle O\rangle = \langle \Omega | O |\Omega\rangle, \label{vev1}
	\end{align}
where $O$ and $|\Omega\rangle$ are the operator and the vacuum in the Heisenberg picture.
The time evolution of $O$ in a general case is complicated.
In the interaction picture, we recast Eq.~\eqref{vev1} into 
\begin{align}
	\langle \Omega |O  |\Omega\rangle & \approx \langle 0 | F^\dagger O_{I}  F
	  |0\rangle,\label{eqHO}
\end{align}
where $|0\rangle$ is the free vacuum, and $F$ is a unitary operator constructed from the free field operators.
We evaluate the two-point function up to the one-loop order by expanding $F$ to third-order in the free fields.
This section provides a brief but self-consistent derivation of Eq.~\eqref{eqHO} and its perturbative expansions. 
One can also find reviews of the in-in formalism in Refs.~\cite{Maldacena:2002vr, Weinberg:2005vy, Chen:2010xka, Wang:2013zva}, and readers who are familiar with the in-in formalism may skip this section.

\medskip
Quantization of classical fields is given by promoting Poisson (Dirac) brackets in the (constraint) canonical formalism to the commutation relations.
Then the canonical equation of motion for a classical field $O$ with respect to a Hamiltonian $H$ becomes the Heisenberg equation for the field operator.
\begin{align}
	\dot {O} = i[H[\phi,\pi;\tau],O]+\frac{\partial O}{\partial \tau},\label{Heiseneq}
\end{align}
where the over-dot is the ordinary derivative with respect to time, $\tau$.
The Hamiltonian operator $H$ is a functional of a canonical field variable $\phi$ and conjugate momentum $\pi$. Using the Hamiltonian density, $\mathcal H$, we have
\begin{align}
	H[\phi,\pi;\tau] = \int d^3x \mathcal H(\phi,\pi; \tau).
\end{align}
In cosmological perturbation theory, $H$ is explicitly dependent on $\tau$ as we consider a time-dependent background. 
Extension to the multi-field operators is straightforward, but here we only consider the single variable for notational simplicity.

If $\partial O/\partial \tau=0$, the time evolution of $O$ is unitary.
Using the time evolution operator $U$ from $\tau_0$ to $\tau$, we get
\begin{align}
	O = U^{-1}O_0U, \label{defOH}
\end{align}
where the subscript $0$ implies the operator evaluated at $\tau = \tau_0$.
Substituting Eqs.~\eqref{defOH} into \eqref{Heiseneq}, we find
\begin{align}
	\dot {O}=
		[-U^{-1}\dot U,O].\label{heiseneq2}
\end{align}
As Eq.~\eqref{heiseneq2} is satisfied for any operators with $\partial O/\partial \tau=0$, Eqs.~\eqref{Heiseneq} and \eqref{heiseneq2} yield 
\begin{align}
	\dot U	=- i U H[\phi,\pi;\tau],\label{HeisenU}
\end{align}
which is different from the expression used in the literature, i.e., one often assumes~(see, e.g., Ref.~\cite{Weinberg:2005vy}.)
\begin{align}
	\dot U	=- i H[\phi,\pi;\tau] U. \label{HeisenU2}
\end{align}
If $\partial H/\partial \tau=0$, we have $H = U^{-1}H U$, so we get Eq.~\eqref{HeisenU2} from Eq.~\eqref{HeisenU}.
However, this is not the case in cosmological perturbation theory.
Indeed, the correct operator ordering of Eq.~\eqref{HeisenU} is used in the following derivation.

If the Hamiltonian is written as $H=\bar H + H_{\rm int}$ with the free-field Hamiltonian $\bar H$ and the interaction part $H_{\rm int}$, the perturbative expansion with respect to the free-field operators is convenient.
We introduce the interaction picture field that evolves as  
\begin{align}
	\dot {O}_{I} = i[\bar H[\phi_{I},\pi_{I};\tau],O_{I}].
\end{align}
The time evolution of $O_I$ is similarly written as
\begin{align}
	O_{I} &= U_I^{-1} O_{I0}U_I,\\
	\dot U_I	&=- i U_I \bar H[\phi_{I},\pi_{I};\tau].\label{intpict}
\end{align}
Assuming $O_{I0} = O_{0}$, the interaction picture field writes the Heisenberg operator as 
\begin{align}
	O =F^{-1} O_I F,~F\equiv U_I^{-1} U.\label{HtoI}
\end{align}
We find the equation of motion for $F$ by using Eqs.~\eqref{HeisenU} and \eqref{intpict}:
\begin{align}
	 \dot F
	&=
	 	 	 - i F H[\phi,\pi;\tau]	+ i  \bar H[\phi_{I},\pi_{I};\tau]  F 	 	 \label{dotF}.	  
\end{align}
Then, 
\begin{align}
	&F H[\phi,\pi;\tau]F^{-1}\notag \\
	 &= U_I^{-1} H[\phi_{0},\pi_{0};\tau] U_I
	=H[\phi_{I},\pi_{I};\tau],
\end{align}
is the Hamiltonian in the interaction picture, and Eq.~\eqref{dotF} yields
\begin{align}
	\dot F &= -i (H[\phi_{I},\pi_{I};\tau]-\bar H[\phi_{I},\pi_{I};\tau])  F
	\notag \\
	&= -i H_{\rm int}[\phi_{I},\pi_{I};\tau]  F.
\end{align}
Given the initial condition $F=1$ for $\tau=\tau_0$, the solution to this equation is formally obtained as 
\begin{align}
	 F= T \exp\left(
	 -i\int^\tau_{\tau_0}d\tau' H_{{\rm int},I}(\tau')
	 \right),
	 \label{Fdef}
\end{align}
where $T$ is the time ordering operator.
Hereafter the functional dependence is suppressed for notational simplicity:
\begin{align}
	H_{{\rm int},I}(\tau') \equiv \int d^3 x \mathcal H_{{\rm int}}\left (\phi_{I}(\tau', \mathbf x),\pi_{I}(\tau', \mathbf x);\tau'\right).
\end{align} 
$\tau$ dependence is also suppressed when it is obvious from the context.

\begin{figure}[hbt]
  \includegraphics[width=0.35\textwidth]{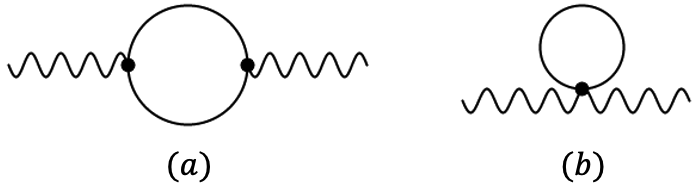}
  \caption{The one-loop order Feynman diagrams considered here: $(a)$ and $(b)$ corresponds to $P_{h2}$ and $P_{h1}$, respectively, in our calculation.}
  \label{fyn}
\end{figure}

\medskip
We obtained the Heisenberg operator written by the interaction picture field in Eq.~\eqref{HtoI}.
The next step is the calculation of the expectation value.
Following the prescription in the in-out formalism, one often shows 
\begin{align}
	\lim_{\tau_0\to -\infty} \langle \Omega | O|\Omega \rangle =\lim_{\tau_0\to -\infty(1-i\epsilon)} \langle 0| O|0 \rangle,\label{eq:vev}
\end{align}
where $\tau_0$ is the initial time of inflation.
We find the above equation as follows.
For simplicity, let us consider that $H$ is time-independent for the $i\epsilon$ prescription and restore the time dependence of $H$ afterward.
Then we find
\begin{align}
		\langle 0|O  |0\rangle  &= \langle 0| e^{i (\tau-\tau^*_0)H} O_{0}  e^{ -i (\tau-\tau_0)H} | 0 \rangle, 
\end{align}
where $O_0$ implies the operator at $\tau=\tau_0$.
Inserting the identity operator expanded by the Hamiltonian eigenstates, $1 = \sum_{n}|\Omega_n \rangle \langle \Omega_n|$ with $H|\Omega_n\rangle = E_{\Omega_n}|\Omega_n \rangle$ and $|\Omega_0\rangle = |\Omega \rangle$, one finds
\begin{align}
\label{ininvc}
	\langle 0|O  |0\rangle &=  \sum_{n,m} e^{-i(\tau-\Re[\tau_0])  ( E_{\Omega_n}- E_{\Omega_m}) -\Im[\tau_0]( E_{\Omega_n}+E_{\Omega_m}) }\notag \\
	&\times  \langle  0|\Omega_m\rangle  \langle \Omega_m | O_{0}|\Omega_n \rangle \langle \Omega_n| 0\rangle.
\end{align}
After the $i\epsilon$ prescription, we get
\begin{align}
	 \lim_{\tau_0\to -\infty(1-i\epsilon)} \langle 0|O  |0\rangle
	= e^{-2\infty \epsilon E_{\Omega_0}} |\langle  0|\Omega \rangle|^2  \langle \Omega | O_0|\Omega \rangle. 
\end{align}
Thus, $E_{\Omega_n}\neq E_{\Omega_m}$ configuration is oscillated away for a $-i \epsilon$ rotated large time interval, and the diagonal components are Boltzmann suppressed by $e^{-2\infty \epsilon E_{\Omega_n}}$.
The normalization factor is determined by considering $O =1$, and one finds $e^{-2\infty \epsilon E_{\Omega_0}} |\langle  0|\Omega \rangle|^2=1$.
Then $\langle \Omega|O  |\Omega\rangle$ is similarly expanded in the limit, and Eq.~\eqref{eq:vev} is shown.
Substituting Eq.~\eqref{Fdef} into Eq.~\eqref{HtoI}, and using Eq.~\eqref{eq:vev}, one finds
\begin{align}
	\langle O\rangle & = \lim_{\tau_0\to -\infty(1-i\epsilon)} \langle 0 | \bar T \exp\left(
	 i\lambda \int^\tau_{\tau_0^*}d\tau' H_{{\rm int},I}(\tau')
	 \right)
	 \notag \\
	 &\times  O_I(\tau)  T \exp\left(
	 -i\lambda \int^\tau_{\tau_0}d\tau'' H_{{\rm int},I}(\tau'')
	 \right) |0\rangle,\label{eq45}
	 \end{align}
where we introduced an order counting parameter $\lambda$, and $\bar T$ is the anti-time ordering operator.
Eq.~\eqref{eq45} is expanded into
\begin{align}
	\langle O\rangle = \sum_{\lambda=0}^\infty  \lambda^n \langle O\rangle _n,\label{Ovev}
\end{align}
where we introduced
\begin{align}
\langle O\rangle_0&=\langle 0|O_{I}(\tau)|0\rangle,	\label{O0}
\\
\langle O\rangle_1&=2\Im  \int^\tau_{\tau_0}d\tau' \langle 0|O_I(\tau) H_{{\rm int},I}(\tau')|0\rangle,\label{O1}
\\
\langle O\rangle_{2a}&=  \int^\tau_{\tau_0^*}d\tau' \int^\tau_{\tau_0}d\tau''  
\notag \\
&\times    \langle 0|  H_{{\rm int},I}(\tau') O_I(\tau) H_{{\rm int},I}(\tau'')	|0\rangle ,\label{O2a}
\\
\langle O\rangle_{2b}&=- 2\Re \int^\tau_{\tau_0}d\tau'\int^{\tau'}_{\tau_0}d\tau'' 
\notag \\
&\times \langle 0| O(\tau)  H_{{\rm int},I}(\tau') H_{{\rm int},I}(\tau'')|0\rangle,\label{O2b}
\end{align}
with $\langle O\rangle_2\equiv \langle O\rangle_{2a}+\langle O\rangle_{2b}$.
In the end of calculation, we set $\tau=0$.
Eqs.~\eqref{O1} to \eqref{O2b} are the basic equations we use in this paper.
Hereafter we set the order counting parameter $\lambda =1$.
$\mathcal O(\lambda^2)$ and $\mathcal O(\lambda^1)$ contributions are diagrammatically represented in Fig.~\ref{fyn}.

\section{Inflationary one-loop calculations}\label{inf-loop}

In the previous section, we reviewed the in-in formalism.
We apply the method to inflationary cosmology in this section.
In cosmological perturbation theory, we expand the full Hamiltonian around the inflationary background.
The second-order Hamiltonian is the free Hamiltonian in this setup, and cosmological perturbations are the interaction picture fields.
The interaction Hamiltonian is the rest of the nonlinear corrections expanded by the linear perturbations.
In this section, we derive the interaction Hamiltonian from the action and find the loop power spectrum.

\subsection{Cosmological setup}
There exists gauge freedom in cosmological perturbation theory as identifying a background spacetime with the physical spacetime is not unique.
We need to fix the gauge for quantization to properly count the dynamical degrees of freedom and thus for the in-in formalism.
In this paper, we consider the uniform curvature gauge for the scalar perturbations where the scalar degrees of freedom are those of matter sectors after solving the constraint equations and the spatial curvature is zero.
As we expand the action to second order in the tensor perturbations, the gauge condition for tensor perturbations should also be specified.
Following Maldacena~\cite{Maldacena:2002vr}, the spatial component of the metric tensor is written as 
\begin{align}
	g_{ij} = a^2 e^{h_{ij}},\label{gijdef}
\end{align}
where $a$ is the background scale factor, and we defined
\begin{align}
	e^{h_{ij}} \equiv \delta_{ij} + h_{ij} +\frac{1}{2}h_{i}{}^{k}h_{kj}+\mathcal O(h^3),\label{ehij}
\end{align}
and the transverse-traceless (TT) condition is given by 
\begin{align}
	\partial^ih_{ij} = \delta^{ij}h_{ij} = 0.
\end{align}
In this paper, the spatial index is raised and lowered by the background spatial metric $\delta^{ij}$ and $\delta_{ij}$.
In the previous literature, one often used 
\begin{align}
	\bar g_{ij} = a^2 (\delta_{ij}+\bar h_{ij}),\label{barhij}
\end{align}
instead of Eq.~\eqref{ehij}, which are equivalent at linear order.
At nonlinear order, the latter tensor gauge condition implies
\begin{align}
	\ln \left( \sqrt{\det |\delta_{ij} + \bar h_{ij} | }\right) =  \frac{1}{4}\bar h_{ij}\bar h^{ij}+\mathcal O(h^3).
\end{align}
Thus, $\bar h_{ij}$ perturbs the volume element, which is the curvature perturbation that should be zero in the uniform curvature gauge.  
Gravitational waves or tensor fluctuations are distortions in spacetime without changing volume.
Hence, it should be introduced such that the volume element is not perturbed by $h_{ij}$.
Eq.~\eqref{ehij} is an example of such a parameterization, i.e., we have
\begin{align}
	\det |e^{ h_{ij}} | =  1.
\end{align}
Therefore, we use Eq.~\eqref{gijdef} instead of Eq.~\eqref{barhij}.
The lapse and shift are slow-roll suppressed compared to $\delta \chi$ in this gauge, so scalar fluctuations are represented only by the matter action to the leading order in the slow-roll parameter~\cite{Maldacena:2002vr}. 
This paper considers quantum corrections of the tensor fluctuations due to a minimally coupled spectator scalar field $\chi$.
With Eq.~\eqref{gijdef}, the possible interaction between $\chi$ and $h_{ij}$ is model-independent and only appears in the kinetic term
\begin{align}
	 S_{\rm full}\supset -\frac{1}{2}\int d^4 x \sqrt{-g}g^{\mu\nu}\nabla_\mu \chi \nabla_\nu \chi,\label{def:kin}
\end{align}
where $S_{\rm full}$ is the full inflationary action.  
We denote the spectator scalar field fluctuation by $\delta \chi$.
Then the relevant part of Eq.~\eqref{def:kin} is written as
\begin{align}
	 \int d\tau a^2 \int d^3x  \left(\frac{1}{2}h^{ij} - \frac{1}{4}h^{ik} h_{k}{}^{j}\right) \partial_i \delta \chi \partial_j \delta \chi.\label{action2}
\end{align}
We read the interaction Lagrangian $L_{\rm int}$ from Eq.~\eqref{action2} and Legendre transform gives $H_{\rm int}=-L_{\rm int}$ as the interaction do not involve time derivative couplings.
Then we find the interaction Hamiltonian
\begin{align}
	H_{\rm int} \equiv a^2 \int d^3x  \left( - \frac{1}{2}h^{ij} + \frac{1}{4}h^{ik} h_{k}{}^{j}\right) \partial_i \delta \chi \partial_j \delta \chi.\label{Ham1}
\end{align}
As far as we consider minimally coupled scalar field, the interaction Hamiltonian is written as Eq.~\eqref{Ham1} for a wide class of theories since potential terms like $\sqrt{-g}V(\chi,\cdots)$ do not contain the tensor fluctuation in the present gauge condition~\eqref{gijdef}.

Fourier integrals of cosmological perturbations are written as
\begin{align}
		\delta  \chi(\tau, \mathbf x)  &= \int \frac{d^3 q}{ (2\pi)^{3}}e^{i\mathbf q\cdot \mathbf x}\delta  \chi_{\mathbf q}(\tau),
		\label{chi:ft}
		\\	
		h_{ij}(\tau, \mathbf x)  &= \int \frac{d^3 q}{ (2\pi)^{3}}e^{i\mathbf q\cdot \mathbf x} \sum_{s=\pm 2}e_{ij}^s(\hat q)h^s_{\mathbf q}(\tau),
		\label{h:ft}
\end{align}
where $\hat q \equiv \mathbf q/|\mathbf q|$. 
The tensor fluctuations are expanded by the polarization tensor $e_{ij}^{s}(\hat q)$, which satisfies
\begin{align}
	e_{ij}^{s}(\hat q)  e^{ij,s'}(\hat q)^* = \delta^{ss'},~e_{ij}^{s}(-\hat q)=e_{ij}^{s}(\hat q)^*, \label{ortho}
\end{align}
where $\delta^{ss'}$ is the Kronecker delta.
Using Eqs.~\eqref{chi:ft} and \eqref{h:ft}, Eq.~\eqref{Ham1} is written as 
\begin{align}
	H_{\rm int} = H^{(3)}_{\rm int} + H^{(4)}_{\rm int},
\end{align}
where Fourier integrals of the interaction Hamiltonians are written as
\begin{align}
	H^{(3)}_{{\rm int}} & =\frac{1}{2}\prod_{A=1}^3\left(\int \frac{d^3p_A}{(2\pi)^{3}}\right)(2\pi)^3\delta\left(\sum_{A=1}^3 \mathbf p_A\right) \sum_{s}  
	\notag \\
	\times &  a^2 h^s_{\mathbf p_1} e^{ij,s}(\hat p_1) p_{2i}p_{3j}  \delta \chi_{\mathbf p_2}  \delta \chi_{\mathbf p_3},
\label{H3:def}
\\
		H^{(4)}_{{\rm int}} 
	&=
	 -\frac{1}{4}\prod_{A=1}^4\left(\int \frac{d^3p_A}{(2\pi)^{3}}\right)(2\pi)^3\delta\left(\sum_{A=1}^4 \mathbf p_A\right)  \sum_{s_1,s_2} 
\notag 	 \\
	 \times  a^2  &e^{ik,s_1}(\hat p_1)e_{k}{}^{j,s_2}(\hat p_2) p_{3i}p_{4j}  h^{s_1}_{\mathbf p_1}h^{s_2}_{\mathbf p_2} \delta \chi_{\mathbf p_3}  \delta \chi_{\mathbf p_4},
\label{H4:def}
\end{align}
where $\delta$ is the 3-dimensional delta function.
As discussed in the previous section, the second-order action describes the free theory in cosmological perturbation theory, and the linear perturbations are the interaction picture fields.
Following the standard quantization procedures, the Fourier transforms of cosmological perturbations are expanded into the creation and annihilation operators as
\begin{align}
	\delta \chi_{\mathbf q}(\tau) & =u_q(\tau)a_{\mathbf q}+ u^*_q(\tau) a^\dagger_{-\mathbf q},\\
	h^s_{\mathbf q}(\tau) & =v_q(\tau) b^s_{\mathbf q}+ v^*_q(\tau) b^{s\dagger}_{-\mathbf q},
\end{align}
where $u_q$ and $v_q$ are the corresponding positive frequency mode functions.
For example, the ground state mode functions for free spectator scalar and tensor fluctuations in general relativity are given as
\begin{align}
	u^{\rm GS}_q(\tau) &= \frac{H}{\sqrt{2q^3}}(1+iq\tau)e^{-iq\tau},\label{defuq}\\
	v^{\rm GS}_q(\tau) &= \frac{2H}{M_{\rm pl}\sqrt{2q^3}}(1+iq\tau)e^{-iq\tau}.\label{defvq}
\end{align}
The annihilation and creation operators of scalar and tensor fluctuations satisfy 
\begin{align}
		a_{\mathbf q}|0\rangle &= b^s_{\mathbf q}|0\rangle = 0,
\end{align}
and the non-vanishing commutation relations are
\begin{align}
	[a_{\mathbf q},a^\dagger_{-\bar{\mathbf q}}] &= (2\pi)^3\delta(\mathbf q+\bar{\mathbf q}),
	\\
	[b^s_{\mathbf q},b^{\bar s,\dagger}_{-\bar{\mathbf q}}] &= (2\pi)^3\delta^{s\bar s}\delta(\mathbf q+\bar{\mathbf q}).
\end{align}

\subsection{Loop calculation}
Given interaction Hamiltonians \eqref{H3:def} and \eqref{H4:def}, we are ready to compute the VEV of
\begin{align}
	O_I (\tau)= \sum_{s=\pm2}h^s_{\mathbf q}(\tau) h^s_{\bar{\mathbf q}}(\tau),
\end{align}
using Eq.~\eqref{eq45}.
We compute the power spectrum $P_h$ defined as
\begin{align}
	\left.	\left\langle \sum_{s=\pm2}h^s_{\mathbf q}(\tau) h^s_{\bar{\mathbf q}}(\tau) \right\rangle \right|_{\tau=0} &=(2\pi)^3 \delta(\mathbf q+\bar{\mathbf q})P_{h}(q),
\end{align}
order by order in $\lambda$ by using Eqs.~\eqref{O0} to \eqref{O2b}.
The total tensor power spectrum, including one-loop correction, is 
\begin{align}
	P_h = P_{h0} + P_{h1} + P_{h2a} + P_{h2b},\label{sumPh}
\end{align}
where the subscript implies the number of vertices in $\lambda$; $P_{hn}=\mathcal O(\lambda^n)$. 
Below we evaluate each term of Eq.~\eqref{sumPh}.

\subsection*{$\mathcal O(\lambda^0)$}
First, we derive the loop correction without any assumptions about the background spacetime or the mode functions.
Calculation of Eq.~\eqref{O0} is straightforward.
From Eq.~\eqref{58.5}, we find the tree-level power spectrum
\begin{align}
	P_{h0}(q) = 2 |v_q(0)|^2. 
\end{align}
This is the power spectrum of the primordial gravitational waves in the standard linear perturbation theory.
For the ground state~\eqref{defvq} at $\tau=0$, we have 
\begin{align}
	P_{h0}(q) = \frac{4H^2}{M_{\rm pl}^2q^3} =\frac{2\pi^2}{q^3}\cdot \frac{2H^2}{\pi^2M_{\rm pl}^2} \label{P11def}.
\end{align}

\subsection*{$\mathcal O(\lambda^1)$}
Next, we compute the diagram of $\mathcal O(\lambda^1)$.
Substituting Eqs.~\eqref{H3:def} and \eqref{H4:def} into Eq.~\eqref{O1}, one finds
\begin{align}
& \left\langle \sum_{s=\pm2}h^s_{\mathbf q}(\tau) h^s_{\bar{\mathbf q}}(\tau) \right\rangle_1 
\notag 
\\
=	&-\frac{1}{2}\Im \int^\tau_{\tau_0}d\tau'  
 \prod_{A=1}^4\left(\int \frac{d^3p_A}{(2\pi)^{3}}\right)(2\pi)^3\delta\left(\sum_{A=1}^4 \mathbf p_A\right)   
\notag 	 \\
	 &\times  \sum_{s,s_1,s_2} a(\tau')^2  e^{ik,s_1}(\hat p_1)e_{k}{}^{j,s_2}(\hat p_2) p_{3i}p_{4j}  	   
\notag \\
&\times 
   \langle 0| h^{s}_{\mathbf q}(\tau)h^{s}_{\bar{\mathbf q}}(\tau)    h^{s_1}_{\mathbf p_1}(\tau')h^{s_2}_{\mathbf p_2}(\tau') |0\rangle \notag \\
   &\times \langle 0|\delta \chi_{\mathbf p_3}(\tau')  \delta \chi_{\mathbf p_4}(\tau') |0\rangle.
\end{align}
Note that the contribution from $H^{(3)}_{\rm int}$ vanishes since the linear tensor fluctuation is Gaussian.
Using Eqs.~\eqref{58} and \eqref{60}, we find
\begin{align}
&\left\langle \sum_{s=\pm2}h^s_{\mathbf q}(\tau) h^s_{\bar{\mathbf q}}(\tau) \right\rangle_1
\notag 
\\
=	&(2\pi)^3\delta\left(\mathbf q+\bar{\mathbf q}\right)\Im \int^\tau_{\tau_0}d\tau'  
  a(\tau')^2 \int \frac{d^3p_3}{(2\pi)^{3}} \sum_{s}   
\notag 	 \\
	 &\times    e^{ik,s}(\hat q)^*e_{k}{}^{j,s}(\hat q)  p_{3i}p_{3j}   
   v_q(\tau)^2   v^*_{q}(\tau')^2 |u_{p_3}(\tau')|^2,
\end{align}
where we dropped a (2-loop) bubble diagram eliminated from our consideration after proper normalization.
Indeed, those diagrams result in zero after the polarization sum due to the angular momentum conservation.
Using 
\begin{align}
	\int \frac{d\hat p_3}{4\pi} \hat p_{3i}\hat p_{3j} = \frac{\delta_{ij}}{3},
\end{align}
and Eq.~\eqref{ortho}, we find
\begin{align}
	\int \frac{d\hat p_3}{4\pi} \sum_{s}  e^{ik,s}(\hat q)^*e_{k}{}^{j,s}(\hat q)  \hat p_{3i}\hat p_{3j}  =\frac{2}{3}
\end{align}
Then we get
\begin{align}
	P_{h1} =& v_q(0)^2 \Im \int^0_{\tau_0}d\tau' a(\tau')^2 v^*_{q}(\tau')^2  
 \int \frac{p^4_3dp_3}{3\pi^{2}}  |u_{p_3}(\tau')|^2.\label{P13}
\end{align}

\subsection*{$\mathcal O(\lambda^2)$}

Lastly, we consider $\mathcal O(\lambda^2)$ contributions that we named $P_{h2a}$ and $P_{h2b}$.
  Using Eqs.~\eqref{O2a} and~\eqref{H3:def}, one obtains
\begin{align}
 &\left\langle \sum_{s=\pm2}h^s_{\mathbf q}(\tau) h^s_{\bar{\mathbf q}}(\tau) \right\rangle_{2a}
  \notag \\
  =&
  \frac{1}{4} \int^\tau_{\tau_0^*}d\tau' a(\tau')^2 \int^\tau_{\tau_0}d\tau''   a(\tau'')^2   \prod_{A=1}^6\left(\int \frac{d^3p_A}{(2\pi)^{3}}\right)
  \notag \\
  &\times (2\pi)^3\delta\left(\sum_{A=1}^3 \mathbf p_A\right)(2\pi)^3\delta\left(\sum_{A=4}^6 \mathbf p_A\right)   
	\notag \\
&	\times \sum_{s,s_1,s_4} e^{ij,s_1}(\hat p_1) p_{2i}p_{3j}    e^{kl,s_4}(\hat p_4) p_{5k}p_{6l} 
	\notag \\
&	\times  \langle 0| h^{s_1}_{\mathbf p_1}(\tau')h^s_{\mathbf q}(\tau)h^s_{\bar{\mathbf q}}(\tau) h^{s_4}_{\mathbf p_4}(\tau'')|0\rangle
\notag \\
&\times \langle 0| \delta \chi_{\mathbf p_2}(\tau')  \delta \chi_{\mathbf p_3} (\tau')  \delta \chi_{\mathbf p_5}(\tau'')  \delta \chi_{\mathbf p_6}(\tau'')	|0\rangle
	\label{63}.
\end{align}
Using Eqs.~\eqref{59}, \eqref{60}, and the index symmetry for $\mathbf p_2$ and $\mathbf p_3$, we find
\begin{align}
 &\left\langle \sum_{s=\pm2}h^s_{\mathbf q}(\tau) h^s_{\bar{\mathbf q}} (\tau)\right\rangle_{2a}
  \notag \\
  =&
  (2\pi)^3\delta\left(\mathbf q+ \bar{\mathbf q}\right) \int^\tau_{\tau_0^*}d\tau' a(\tau')^2  \int^\tau_{\tau_0}d\tau''    a(\tau'')^2 
  \notag \\
  &\times \prod_{A=2}^3\left(\int \frac{d^3p_A}{(2\pi)^{3}}\right) (2\pi)^3\delta\left(\sum_{A=2}^3 \mathbf p_A-\mathbf q\right)   
	\notag \\
&	\times \sum_{s} e^{ij,s}(\hat q) p_{2i}p_{3j}    e^{kl,s}(\hat q)^* p_{2k}p_{3l} v_{q}(\tau) v^*_{q}(\tau)    
	\notag \\
&	\times v_{q}(\tau') u_{p_2}(\tau')u_{p_3}(\tau') v^*_{q}(\tau'') u^*_{p_2}(\tau'')u^*_{p_3}(\tau'')
	\label{64},
\end{align}
where we dropped bubble graphs, and there is no contribution from tadpole graphs.
We can simplify the angular integrals in the above equation by using Appendix~\ref{angular_int}.
Combining Eqs.~\eqref{sinsin}, \eqref{sinsin2}, and \eqref{wakeru}, one finds
\begin{align}
	& \prod_{A=2}^3\left(\int \frac{d^3p_A}{(2\pi)^{3}}\right)(2\pi)^3\delta\left(\sum_{A=2}^3 \mathbf p_A - \mathbf q\right)  
	\notag \\
	&\times  \sum_{s=\pm2} |e^{ij,s}(\hat q) p_{2i}p_{3j}    |^2 f(p_2,p_3) 
	\notag \\
	=&   \int_0^{\infty} d p_2 \int_{|p_2-q|}^{p_2+q} dp_3  \bar w(q; p_2,p_3)f(p_2,p_3),\label{sekibunhenkan}
\end{align}
where we defined
\begin{align}
	&\bar w(q; p_2,p_3) \notag \\
	&\equiv  \frac{p_2p_3 \left(p_2^4-2 p_2^2 \left(p_3^2+q^2\right)+\left(p_3^2-q^2\right)^2\right)^2}{128\pi^2q^5}\label{defbarw}.
\end{align}
Using Eq.~\eqref{sekibunhenkan}, Eq.~\eqref{64} simplifies to 
\begin{align}
 P_{h2a}
  =&
 |v_{q}(0)|^2 \int_0^{\infty} d p_2 \int_{|p_2-q|}^{p_2+q} dp_3  \bar w(q; p_2,p_3) \notag \\
 &\times   \left| \int^0_{\tau_0^*}d\tau'     a(\tau')^2     v_{q}(\tau') u_{p_2}(\tau')u_{p_3}(\tau')\right|^2 
	.\label{p22a}
\end{align}
$P_{h2b}$ can be calculated similarly.
Using Eqs.~\eqref{59} and \eqref{60} for Eqs.~\eqref{O2b} and~\eqref{H3:def}, and dropping the bubble graphs, we find 
\begin{align}
	P_{h2b} &=
	 -2v_q(0)^2 \Re \int_0^{\infty} d p_2 \int_{|p_2-q|}^{p_2+q} dp_3  \bar w(q; p_2,p_3)  
	 \notag \\
&\times   
 \int^0_{\tau_0}d\tau'a(\tau')^2 v^*_{q}(\tau') u_{p_2}(\tau') u_{p_3}(\tau') 
 \notag \\
 &\times \int^{\tau'}_{\tau_0}d\tau''  a(\tau'')^2 v^*_{q}(\tau'') u^*_{p_2}(\tau'') u^*_{p_3}(\tau'').
 \label{p22b}
\end{align}
Note that $\Re$ and $\Im$ imply that we take the real and imaginary parts of the whole equations, including the integral domain.
This section derived general one-loop formulas without specifying mode functions of scalar and tensor perturbations.
Eqs.~\eqref{P13}, \eqref{p22a} and \eqref{p22b} will be useful for various purposes.

\section{One-loop corrections from excited states}
\label{loopfromex}
This section considers the loop corrections from an excited scalar field at some specific momenta.
Such a setup is interesting mainly for two reasons.
Firstly, the loop integral is simplified.
We do not have to consider the UV or IR divergence issue since the loop integral is evaluated for specific momenta whose physics is completely determined by a given model.
Secondly, such a scale-dependent scalar field is phenomenologically considered in the context of primordial black hole formation.
The loop corrections from enhanced scalar fields may be observationally testable.
Let us derive the one-loop formulas for monotonically excited scalar fields in this section.

\subsection{One-loop formulas for sharp excited states}
In the previous section, we saw that $\mathcal O(\lambda^2)$ contributions \eqref{p22a} and \eqref{p22b} take the following form
\begin{align}
	P_{h2} =  \int_0^{\infty} d p_2 \int_{|p_2-q|}^{p_2+q} dp_3  \bar w(q; p_2,p_3) f(p_2,p_3).\label{67}
\end{align}
The simplest situation is that we only have the scalar field excitation for $p_2=p_3=p_*$ modes. 
In that case, the evaluation of Eq.~\eqref{P13} is straightforward, and the integrand in Eq.~\eqref{67} should be replaced as
\begin{align}
	f^\delta(p_2,p_3) =\delta\left(\ln (p_2/p_*) \right)\delta\left(\ln (p_3/ p_*) \right) f(p_*,p_*).\label{deltaintro}
\end{align}
Using Eq.~\eqref{A6}, we obtain
\begin{align}
	P_{h2}   &=      \frac{p_*^4(4p_*-q^2)^2}{128\pi^2 q}\Theta_{2-q/p_*}f(p_*,p_*),\label{eqint}
	   \end{align}	   
where $\Theta_{2-q/p_*}$ is the Heaviside step function of $2-q/p_*$.
The Heaviside step function implies the momentum conservation upon generating tensor fluctuations from the scalar fluctuations.
	   On the other hand, the loop integral in Eq.~\eqref{P13} is straightforward.
Using Eq.~\eqref{eqint} for Eqs.~\eqref{p22a} and \eqref{p22b}, we obtain
\begin{align}
	\frac{P^{\delta}_{h1}}{P_{h0}} =&\frac{p^5_*}{6\pi^{2}} \Im \int^0_{\tau_0}d\tau'  a(\tau')^2 v^*_{q}(\tau')^2  
    |u_{p_*}(\tau')|^2,\label{P13ln}
\\
 \frac{P^{\delta}_{h2a}}{P_{h0}}
  =&
\frac{p_*^4(4p_*^2-q^2)^2}{256\pi^2 q}\Theta_{2-q/p_*}  
\notag \\
&\times  \left| \int^0_{\tau_0^*}d\tau'     a(\tau')^2     v_{q}(\tau') u_{p_*}(\tau')^2\right|^2 
	\label{p22aln},
\\
	\frac{P^{\delta}_{h2b}}{P_{h0}} =&
	 - \frac{p_*^4(4p_*^2-q^2)^2\Theta_{2-q/p_*}}{128\pi^2q}  
	 \notag \\
&\times 
\Re \int^0_{\tau_0}d\tau'a(\tau')^2  v^*_{q}(\tau') u_{p_*}(\tau')^2 
\notag \\
&\times 
 \int^{\tau'}_{\tau_0}d\tau'' a(\tau'')^2   v^*_{q}(\tau'') u^*_{p_*}(\tau'')^2.\label{p22bln}
\end{align}
	   
The delta functions in Eq.~\eqref{deltaintro} are written as the superposition of plane waves at all scales: 
\begin{align}
	\delta(\ln \tilde q) =\frac{1}{2\pi} \int_{-\infty}^\infty e^{i(\ln \tilde q) y}dy. \label{deltadef}
\end{align}
From the causality perspective, a wave packet in real space should be localized around a finite region.
Hence, employing Eq.~\eqref{deltadef} implicitly violates causality.
To localize the real space field configuration, we introduce a Gaussian window for the Fourier integral \eqref{deltadef}.
Then we find  
\begin{align}
	\frac{1}{2\pi}  \int_{-\infty}^\infty dy e^{-\frac{1}{2}\Delta^2 y^2} e^{i(\ln \tilde q)y} = \frac{e^{-\frac{[\ln \tilde q ]^2}{2\Delta^2}}}{\sqrt{2\pi}\Delta}.
\end{align}
Thus, more realistic momentum distribution could be written by a lognormal distribution with a finite width $\Delta$ around $p_2=p_3=p_*$.
In this case, we generalize Eq.~\eqref{deltaintro} to
\begin{align}
	f^{\rm LN}(p_2,p_3) =\frac{e^{-\frac{[\ln (p_2/p_*)]^2+[\ln (p_3/p_*)]^2}{2\Delta^2}}}{2\pi \Delta^2} f(p_*,p_*).\label{fln}
\end{align}
In the $\Delta\to 0$ limit, the lognormal distribution recovers the delta function.
For the narrow lognormal peak, the momentum integral is approximated by the peak value at $p_2=p_3=p_*$.
Then Eq.~\eqref{P13} has the same form as Eq.~\eqref{P13ln}.
For $P_{h2}$, we can integrate out one of the momenta trivially, and the remaining integral gives a generalized step function 
\begin{align}
	\Theta^\Delta_{2-q/p_*} \equiv  \frac{e^{2 \Delta ^2}}{2}\left[
\text{erf}\left(\frac{2 \Delta ^2-\ln \left(| 1-q/p_*| \right)}{\sqrt{2} \Delta }\right)
\right.
\notag \\
\left.-\text{erf}\left(\frac{2 \Delta ^2-\ln \left(1+q/p_*\right)}{\sqrt{2} \Delta }
\right)\right]. 
\end{align}
Thus, the Heaviside step function in Eq.~\eqref{eqint} is generalized to $\Theta^\Delta$, as we have
\begin{align}
	\lim_{\Delta\to 0}\Theta^\Delta_{2- q/p_*} = \Theta_{2- q/p_*}.
\end{align}
For a small $q/p_*$, one finds
\begin{align}
	\Theta^\Delta_{2p_*-\tilde q} =\sqrt{\frac{2}{\pi}}\frac{q}{p_*\Delta} + \mathcal O(q^3/p_*^3).\label{Deltatheta}
\end{align}
Thus $\Theta^\Delta$ introduces additional $q/p_*$ factor for the log-normal case.
Hence, $P^{\rm LN}_{h2a/h2b}$ are obtained by replacing $\Theta$ with $\Theta^\Delta$ in $P^{\delta}_{h2a/h2b}$.

\subsection{A model}
\label{model:sec}
We derived the one-loop formulas for the partly enhanced scalar fields in the previous sections.
This section considers a specific model of excited scalar fields and evaluates the loop corrections more concretely.
First, let us consider an enhancement of spectator scalar fluctuation modeled by
\begin{align}
	\delta \chi = e^{\mu H (t_f-t_i)} \delta \chi^{\rm GS},\label{expamp}
\end{align}
where superscript ``GS'' implies the ground state, $\mu$ is a dimensionless time constant, $t_i$ and $t_f$ are the physical time of the initial and final time of the scalar enhancement.
Note that the implementation of spectator field enhancement is not unique. 
We also discuss the spectator scalar field amplification via the Bogoliubov transformation in Section~\ref{BTsec}.
In this paper, we do not specify the enhancement process to amplify the scalar mode but consider the consequences of Eq.~\eqref{expamp}.
Using the conformal time, the enhancement factor in Eq.\eqref{expamp} can be recast into 
\begin{align}
	e^{\mu H (t_f-t_i)} = \left(\frac{\tau_i}{\tau_f}\right)^{\mu},\label{expamp_conf}
\end{align} 
where $\tau_i$ and $\tau_f$ are the conformal time counterpart of $t_i$ and $t_f$.
For simplicity, the scalar fluctuation is in the ground state before the amplitude enhancement, and the enhancement factor is constant for $\tau>\tau_f$.
All details, including the time dependence of the enhancement factor, are contained in 
\begin{align}
	\Xi(\tau) = 
	\begin{cases}
	0,&(\tau<\tau_i),\\
	\left(\frac{\tau_i}{\tau}\right)^\mu,&(\tau_i<\tau<\tau_f), \\
	\left(\frac{\tau_i}{\tau_f}\right)^\mu,&(\tau_f<\tau),	
	\end{cases}
	\label{defxi}
\end{align}
where $\Xi=0$ for $\tau<\tau_i$ implies that we subtracted the vacuum contribution.
We can change the time dependence of $\Xi$ for different models.
Time dependence at $\tau_f<\tau<0$ relies on a realization of how the scalar amplitude is enhanced. 
More realistic enhancement factors could be oscillating or decaying after $\tau=\tau_f$.

Let us consider that the enhancement happens only for $ p= p_*$ modes.
In this case, Eq.~\eqref{expamp} for $p=p_*$ is realized by simply multiplying $\Xi$ and the ground state mode function as
\begin{align}
	u_{p_*} = \Xi u^{\rm GS}_{p_*},\label{udef}
\end{align}
and the tensor fluctuation is not changed: $v_q = v_q^{\rm GS}$.
Eq.~\eqref{udef} could be realized when, e.g.,  the sound speed of the spectator field is exponentially suppressed.
We may also consider such an exponential factor in the parametric resonance. 

We can straightforwardly use Eqs.~\eqref{P13ln} to \eqref{p22bln} for Eq.~\eqref{udef}, and we find
\begin{align}
	 \frac{P^{\delta}_{h1}}{P_{h0}}=&  \frac{H^2}{M_{\rm pl}^2}   
	 \Im  \int^0_{x_0} dx X_{\tilde q}(x),\label{P13delta}
\\
		\frac{P^{\delta}_{h2a}}{P_{h0}}=& \frac{1}{2}\frac{H^2}{M_{\rm pl}^2} \Theta^\Delta_{2-\tilde q} 
 \left| \int^{0}_{x_0} Y_{\tilde q} ( x) dx
		     \right|^2,\label{P22adelta}
\\
		\frac{P^{\delta}_{h2b}}{P_{h0}}=&- \frac{H^2}{M_{\rm pl}^2}
		\Theta^\Delta_{2-\tilde q}
		    \Re \int^{0}_{x_0} dx  \int^{x}_{x_0} dx'  
Y^*_{-\tilde q}(x)Y_{\tilde q}(x'),\label{P22bdelta}
\end{align}	
where $x_a\equiv p_*\tau_a$ ($a=0,i,f$) and $\tilde q\equiv q/p_*$, $a=-1/(H\tau)$ and we introduced
\begin{align}
	X_{\tilde q}(x) &\equiv   \frac{(1+ x^2)(1-i\tilde qx )^2}{6\pi^2 \tilde q^3x^{2}}e^{2i\tilde qx} \Xi(x/p_*),
	\\
	Y_{\tilde q}(x) &\equiv  \frac{(4-\tilde q^2)(1-i\tilde qx)(1-ix)^2}{16\pi \tilde q^2 x^{2}}
	e^{i(\tilde q+2)x} \Xi(x/p_*).
\end{align}
The IR behavior of $P_{h1}$ can be found analytically.
Expanding Eq.~\eqref{P13delta} with respect to $\tilde q$, 
\begin{align}
	\frac{P_{h1}}{P_{h0}} &= -\frac{H^2}{18 \pi^2M_{\rm pl}^2} \left(  \frac{x_f^4}{\mu-2}+ \frac{x_f^2}{\mu-1}\right) \left(\frac{x_i}{x_f}\right)^{2 \mu }
	\notag 
	\\
	&
	+\frac{H^2}{18 \pi^2 M_{\rm pl}^2} \left(\frac{x_i^4}{\mu-2}+ \frac{x_i^2}{\mu-1}\right) + \mathcal O(\tilde q).
\end{align}
For $x_i<x_f<0$ and $\mu\geq 0$, we find
\begin{align}
	\lim_{q/p_*\to 0 }\frac{P_{h1}}{P_{h0}}=-\mathcal O(\Xi^2)<0.\label{P13neg}
\end{align}
Thus, $P_{h1}$ is scale-invariant and negative, i.e., the super-horizon primordial gravitational waves will be reduced by this effect. 
The IR contribution of $P^\delta_{h2}$ is given as
\begin{align}
	\frac{P^{\delta}_{h2}}{P_{h0}}
	=\frac{H^2\Gamma[\Xi]}{24 \pi^{2} M_{\rm pl}^2 \tilde q }	 +\mathcal O(\tilde q),
\label{deltaPint}
\end{align}
where we defined a functional of $\Xi$ as
\begin{align}
&\Gamma[\Xi]=	\int^{0}_{x_i} dx' \int^{x'}_{x_i} dx''
 \frac{x'}{x''^2} \Xi(x'/p_*)^2\Xi(x''/p_*)^2 
 \notag 
 \\
 &\times 
  \left[\left(x'^2+ x''^2-4 x' x''-x'^2x''^2-1\right) \sin (2x'-2x'')\right.
  \notag
  \\
  &
  \left.+2 \left(x'^2 x''-x' x''^2+x'-x''\right) \cos (2 x'-2 x'')\right],
\end{align}
and we used the following relation for $P_{h2a}$:
\begin{align}
	 &\int_{x_0}^{x}dx' \int_{x_0}^{x}dx'' A(x',x'')  
	 \notag \\
	 &=\int_{x_0}^{x}dx'\int_{x_0}^{x'}dx'' 
	\left[ A(x',x'')+A(x'',x')\right].
\end{align}
The red scale dependence of Eq.~\eqref{deltaPint} comes from the fact that the monotonic configuration in the Fourier space violates causality in real space.
Properly accounting for the causality by generalizing the delta-function-like distribution, one finds an additional $\tilde q$ factor from Eq.~\eqref{Deltatheta}, which yields
\begin{align}
	\frac{P^{\rm LN}_{h2}}{P_{h0}}
	=\sqrt{\frac{2}{\pi}}\frac{H^2\Gamma[\Xi]}{24 \pi^{2} M_{\rm pl}^2\Delta}	 +\mathcal O(\tilde q^2).
\label{lnPint}
\end{align}
Thus, we obtain the scale-invariant correction of $\mathcal O(\Xi^4)$.
$P_{h1}=\mathcal O(\Xi^2)$, so the one-loop spectrum is dominated by $P_{h2}$.

\begin{figure*}
  \includegraphics[width=\linewidth]{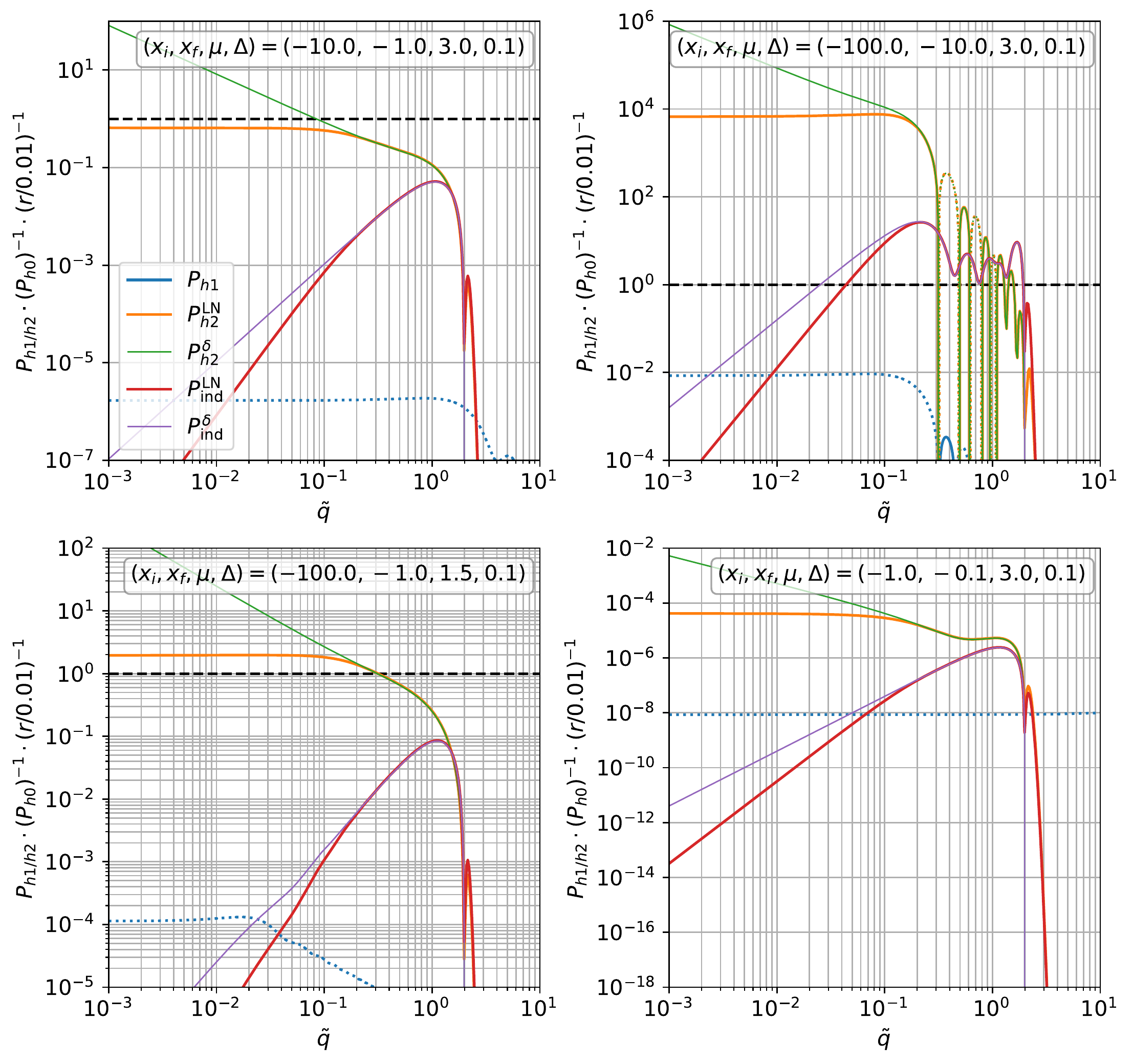}
  \caption{The one-loop corrections divided by the linear tensor power spectrum in units of $r/0.01$ with the linear tensor-to-scalar ratio $r$ are presented. We evaluate the spectra at $\tau=0$. The horizontal axis is the external Fourier momentum $q$ divided by the scalar field peak momentum $p_*$, i.e., $\tilde q \equiv q/p_*$. The black dashed lines imply unity, and the spectra above unity suggest that the loop contributions are dominant over the linear tensor power spectrum. The solid blue, orange, green, red and purple curves represent $P_{h1}$, $P^{\rm LN}_{2}$, $P^{\delta}_{2}$, $P^{\rm LN}_{\rm ind.}$, and $P^{\delta}_{\rm ind.}$, respectively. The dotted curves mean the negative part.
  $x_i\equiv p_* \tau_i$ and $x_f\equiv p_*\tau_f$ are the dimensionless time at the initial and final time of enhancement, $\mu$ is the time constant of the exponential amplification, and $\Delta$ is the width of the lognormal distribution. }
  \label{pow}	
\end{figure*}

Further quantitative details are discussed numerically. 
$\Xi$ in Eq.~\eqref{defxi} is parameterized by 3 independent parameters: the dimensionless initial time $x_i=p_*\tau_i$, and final time of amplification $x_f=p_*\tau_f$, and the dimensionless time constant $\mu$.
Fixing the final enhancement factor as $(x_i/x_f)^\mu = 10^3$, we may consider the following physically different situations:

\begin{enumerate}
\renewcommand{\labelenumi}{(\Alph{enumi})}
	\item Near horizon contribution, $x_f=-1$, $x_i=-10$, $\mu=3$: the enhancement of $\delta \chi$ happens and stops \textit{just before} the horizon exit of $p=p_*$ modes.
	\item Subhorizon contribution, $x_f=-10$, $x_i=-100$, $\mu=3$: the enhancement of $\delta \chi$ happens and stops \textit{well in advance} of the horizon exit of $p=p_*$ modes. 
	\item Sub-to-near horizon contribution, $x_f=-1$, $x_i=-100$, $\mu=1.5$: the enhancement of $\delta \chi$ happens \textit{well in advance} and stops \textit{just before} the horizon exit of $p=p_*$ modes.  
	\item Super-horizon contribution, $x_f=-0.1$, $x_i=-1$, $\mu=3$: the enhancement of $\delta \chi$ happens at super-horizon scale.  
\end{enumerate}

Another unknown parameter is the inflationary Hubble parameter $H$, which normalizes the linear tensor spectrum.
We control the parameter by the \textit{linear} tensor-to-scalar ratio defined as
\begin{align}
	r \equiv  \frac{P_{h0}}{P_{\zeta0}},
\end{align}
where we used Eq.~\eqref{P11def} and $P_{\zeta0}$ is the linear power spectrum of the adiabatic perturbation $\zeta$.
From the CMB measurements, we have the almost scale-invariant scalar spectrum. At the pivot scale $q_p=0.05{h/\rm Mpc}$, we have $q_p^3P_{\zeta0}(q_p)/2\pi^2=2.196\times 10^{-9}$, which yields
\begin{align}
	\frac{H^2}{M_{\rm pl}^2} = \frac{\pi^2}{2}\cdot r \cdot \frac{q^3P_{\zeta0}}{2\pi^2} \approx \left(\frac{r}{0.01}\right)\times 10^{-10}.
\end{align}
As shown in Eqs.~\eqref{P13delta} to \eqref{P22bdelta}, power of $H^2/M_{\rm pl}^2$ counts the number of loops.

In Fig.~\ref{pow}, we show the results of numerical calculations for the above (A) to (D).
We present the one-loop corrections normalized by $P_{h0}$ in units of $r/0.01$.
Hence, the unity in the vertical axis implies that the loop contribution is comparable to the linear tensor fluctuations for $r=0.01$.
The subhorizon spectator field enhances the super-horizon tensor fluctuations from (A) to (C).
On the other hand, the enhancement from the super-horizon scalar field in (D) is relatively suppressed because of causality.
The figure shows that the enhancement of the scalar amplitude at a shorter scale introduces larger one-loop corrections because $\delta \chi$ continuously enhances the tensor fluctuation until $p\sim p_*$ modes exit the horizon.
We see that the IR behavior discussed analytically is reproduced in numerical calculation.
The size of $P_{h1}$ and $P_{h2}$ are loosely related as $P_{h2}\approx \Xi^2 P_{h1}$ near $\tilde q=1$ as the former involves two additional scalar field operators.
We also plot the induced tensor power spectra discussed in the next section.
As discussed later, the induced components are included in the in-in calculation, which is dominant only for subhorizon scales.

\begin{figure*}
  \includegraphics[width=\linewidth]{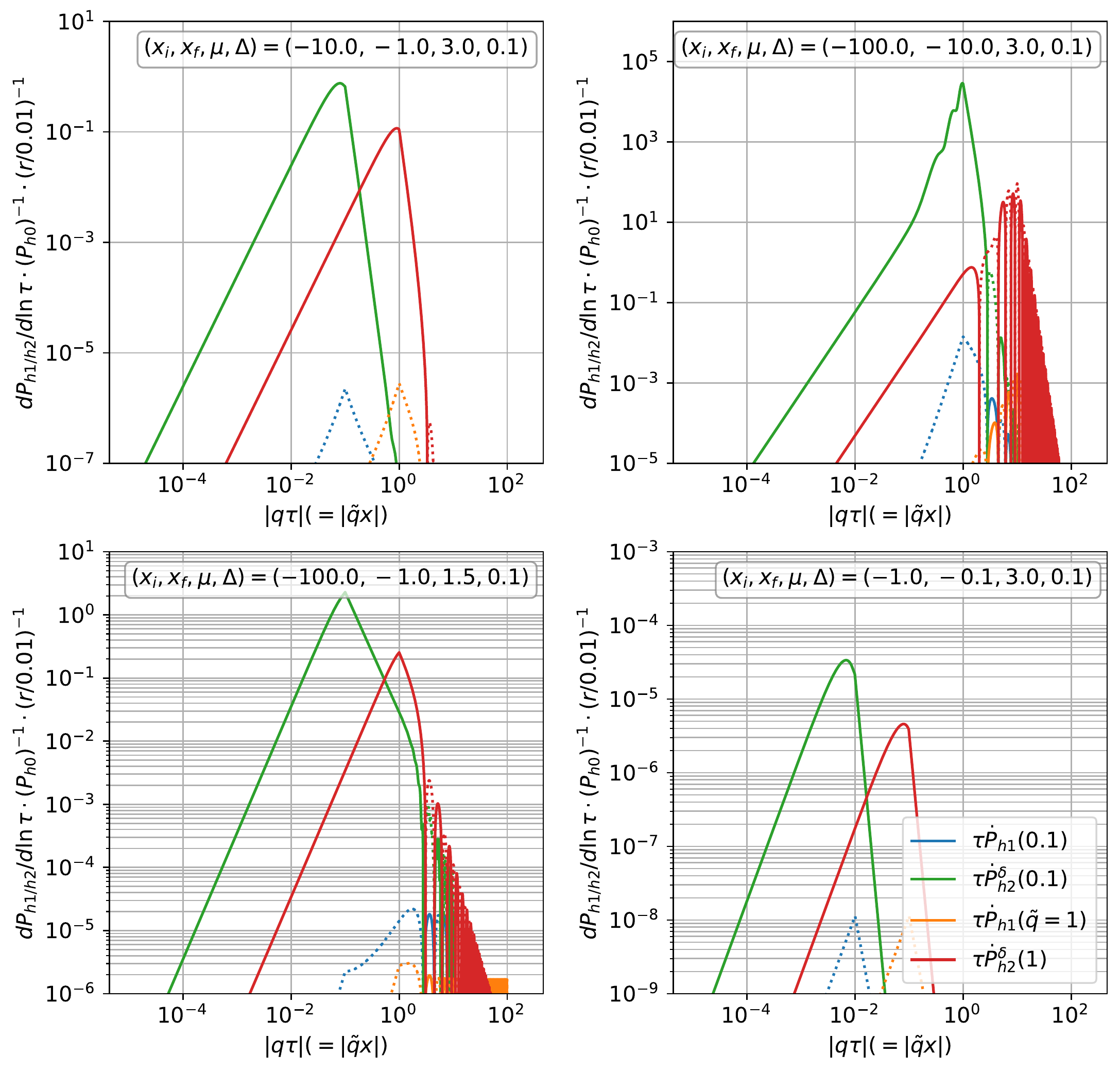}
  \caption{Time evolution of the one-loop spectrum in a cosmic time, $\tau \partial P^\text{1-loop}/\partial \tau$ multiplied by $(P^{\rm tree}_{h})^{-1}\cdot (r/0.01)^{-1}$ with the linear tensor-to-scalar ratio $r$.
 	The horizontal axis is the magnitude of time normalized by the wavenumber of the tensor mode, and the time arrow is from right to left during inflation. 
 	Curves are drawn for the fixed Fourier modes of $\tilde q\equiv q/p_*=1$, and $0.1$, where $q$ is the wavenumber of the selected Fourier modes $p_*$ is that of the peak location.
 	The scalar source and the tensor mode exit the horizon simultaneously for $\tilde q=1$, whereas the tensor mode exit the horizon prior to the scalar source for $\tilde q=0.1$.
 	The orange~(blue) and red~(green) curves represent $\tau \partial P_{h1}/\partial \tau$ and $\tau \partial P^{\delta}_{2}/\partial \tau$ for $\tilde q=1~(0.1)$, respectively.  
 	Each plot indicates that the evolution of the loop correction lasts until the exponential amplification stops, regardless of the scale of the tensor mode.  
 	Dotted curves represent the negative parts.
 	}
  \label{dpow}	
\end{figure*}

\begin{figure*}
  \includegraphics[width=\linewidth]{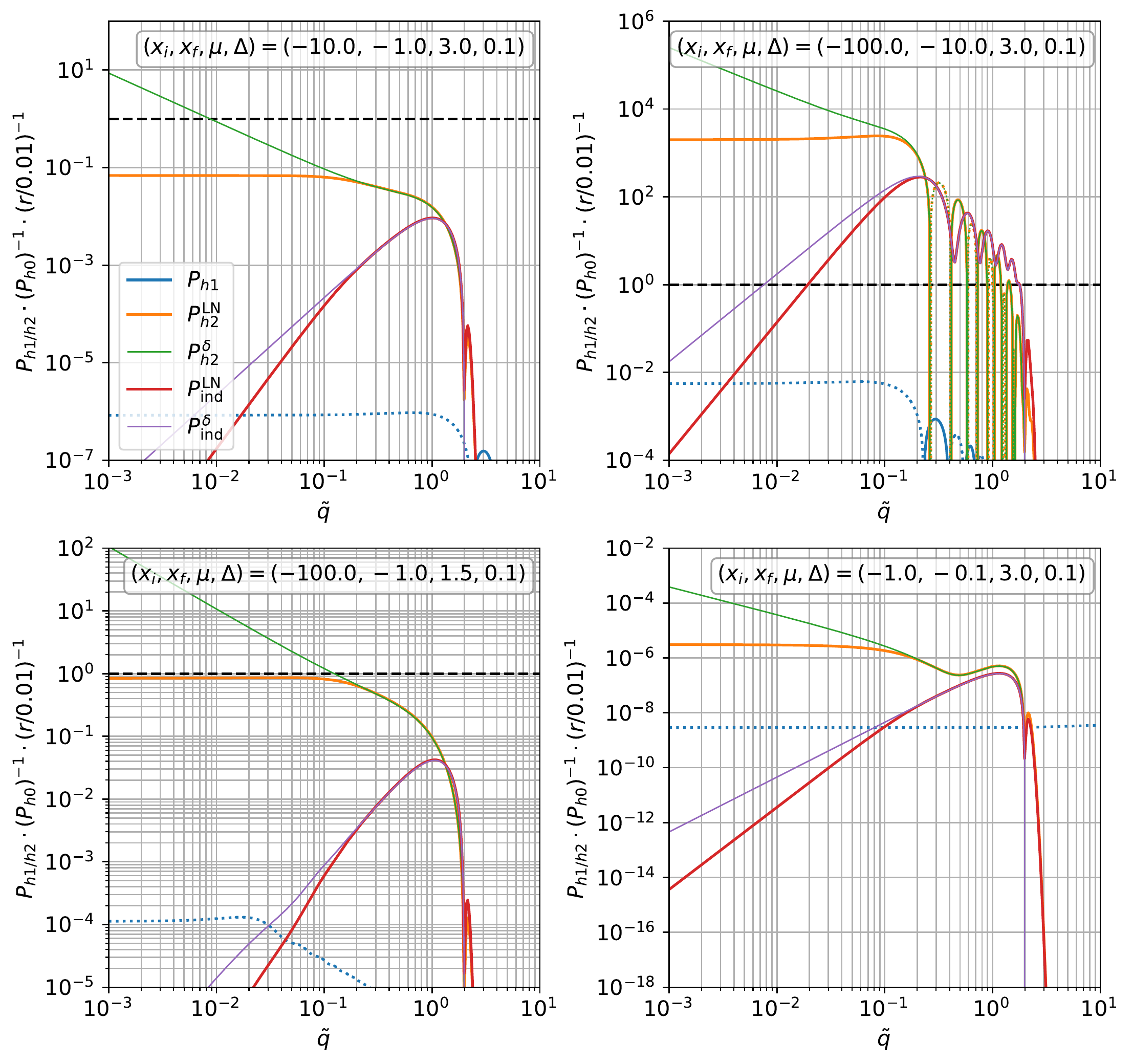}
  \caption{One-loop spectrum when we switch off the scalar field amplification after the enhancement, that is, $\Xi=0$ for $\tau>\tau_f$. 
  Definitions of symbols are the same as Fig.~\ref{pow}.}
  \label{cut_fig}	
\end{figure*}

Several Fourier modes' time evolution is presented in Fig.~\ref{dpow}.
The figure shows that the nonlinear tensor fluctuations' growth~(decay) rate quickly converges to zero after the scalar field amplification stops.
After the exponential enhancement, we keep the constant spectator scalar field, which introduces the loop correction for $\tau>\tau_f$.
The late time evolution could be the dominant component of the final spectrum. At the same time, the IR scaling is irrespective of the time dependence of the enhancement mechanism of the scalar amplitude.
In Fig.~\ref{cut_fig}, we computed the loop correction when we set $\Xi=0$ for $\tau>\tau_f$, that is, we subtracted the late time contribution.
The early time contribution is at most $\mathcal O(10)$\% of the total correction for the present parameters.
Thus, the final spectrum is sensitive to the implementation of the enhancement of scalar amplitude.

\section{Bogoliubov perspective}\label{BTsec}

In the previous section, we saw the super-horizon reduction and enhancement of the tensor power spectrum.
One usually believes that the super-horizon tensor fluctuation variation is prohibited from causality.
Therefore, our results look counterintuitive at first glance.
Indeed, the GW production due to physical processes must respect causality, and the induced GWs are produced only on the subhorizon scale. 
This section explains that the super-horizon correction is not considered as  ``induced GWs'' because the tensor fluctuations are not produced from ``zero''.

\subsection{Super-horizon evolution as a Bogoliubov transformation}
  
We may understand the physical implication of the effect more clearly by inspecting the effect at the field level.  
We can see the field evolution by directly solving the Heisenberg equation.
The Heisenberg operator is written by the interaction picture field as
\begin{align}
	h^s_{\mathbf q,{\rm H}} =& F(\tau;\tau_i)^\dagger h^s_{\mathbf q} F(\tau;\tau_i), \label{Heisen-hop}
\end{align}
where $h^s_{\mathbf q}$ is the interaction picture tensor fluctuation.
The time evolution operators are given as
\begin{align}
	&F(\tau;\tau_i) = 1 - i \lambda \int^\tau_{\tau_i} d\tau' H_{{\rm int},I}(\tau')
	\notag \\
	&
	-\lambda ^2\int^\tau_{\tau_i} d\tau' \int^{\tau'}_{\tau_i} d\tau'' H_{{\rm int},I}(\tau')H_{{\rm int},I}(\tau'')+\mathcal O(\lambda^3),\label{defFop}
\end{align}
where we temporarily restored the order counting parameter $\lambda$.
As $|\tau_i|\ll |\tau_0|$, we may consider $\tau_i=\tau_i^*$.
In this case, substituting Eqs.~\eqref{defFop} into \eqref{Heisen-hop}, we find~\cite{Weinberg:2005vy}
\begin{align}
	&h^s_{\mathbf q,{\rm H}}(\tau) =h^s_{\mathbf q}(\tau) +  i \lambda \int^\tau_{\tau_i} d\tau' [H_{{\rm int},I}(\tau'),h^s_{\mathbf q}(\tau)]
	\notag \\
	&
-\lambda^2 \int^\tau_{\tau_i} d\tau' \int^{\tau'}_{\tau_i} d\tau'' [H_{{\rm int},I}(\tau''),[H_{{\rm int},I}(\tau'),h^s_{\mathbf q}]]
+\mathcal O(\lambda^3).\label{YFeq}
\end{align}
Eq.~\eqref{YFeq} is regarded as a perturbative solution to the quantum equation of motion~\eqref{Heiseneq}.
Since we have $F^\dagger h^2 F =(F^\dagger h F)(F^\dagger h F),$ Eq.~\eqref{YFeq} should reproduce the loop spectrum obtained in the in-in formalism.
The one-loop spectrum contributions from $\mathcal O(\lambda^2)$ in Eq.~\eqref{YFeq} can be understood as quantum correction for the first iterative component. 
$\mathcal O(\lambda^1)$ in Eq.~\eqref{YFeq} are understood as the Born approximation effect.

\medskip
$\mathcal O(\lambda^1)$ corrections are composed of two terms.
The first term comes from the third-order interaction Hamiltonian:
\begin{align}
	&i\int^\tau_{\tau_i} d\tau' [H^{(3)}_{{\rm int},I}(\tau'),h^s_{\mathbf q}(\tau)] 
\notag \\
= &\frac{i}{2} \int^\tau_{\tau_i} d\tau' \prod_{A=1}^3\left(\int \frac{d^3p_A}{(2\pi)^{3}}\right)(2\pi)^3 \delta\left(\sum_{A=1}^3 \mathbf p_A\right)   \sum_{s_1}
	\notag \\
	\times &  a(\tau')^2 [h^{s_1}_{\mathbf p_1}(\tau'),h^s_{\mathbf q}(\tau)] e^{s_1}_{ij}(\hat p_1) p_{2i}p_{3j}  \delta \chi_{\mathbf p_2}(\tau')  \delta \chi_{\mathbf p_3}(\tau').
	\label{ind1}
\end{align}
Simplifying the commutator of the tensor fluctuations as
\begin{align}
	&[h^{s_1}_{\mathbf p_1}(\tau'),h^s_{\mathbf q}(\tau)]
\notag
\\
=&( v_{q}(\tau')v^*_{q}(\tau)-v^*_{q}(\tau')v_{q}(\tau))(2\pi)^3\delta(\mathbf q+\mathbf p_1)\delta^{ss_1},
\end{align}
Eq.~\eqref{ind1} yields
\begin{align}
	&i\int^\tau_{\tau_i} d\tau' [H^{(3)}_{{\rm int},I}(\tau'),h^s_{\mathbf q}(\tau)] =  \int^\tau_{\tau_i} d\tau' G_q(\tau;\tau') S^s_{\mathbf q}[\delta \chi(\tau')] \label{semi3}.
\end{align}
where we defined
\begin{align}
	G_q(\tau;\tau') &= \frac{v_{q}(\tau')v^*_{q}(\tau) - v^*_{q}(\tau')v_{q}(\tau)}{W(\tau')} ,\label{Green}\\
	W(\tau) & = \frac{i}{ a(\tau)^2},\label{Wronskiandef}\\
	S^s_{\mathbf q}[\delta \chi]& = -\frac{1}{2}\prod_{A=2}^3\left(\int \frac{d^3p_A}{(2\pi)^{3}}\right)\delta\left(\mathbf q - \sum_{A=2}^3 \mathbf p_A\right)  
	\notag \\
	&\times   e^{ij,s,*}(\hat q) p_{2i}p_{3j}  \delta \chi_{\mathbf p_2}  \delta \chi_{\mathbf p_3}. \label{source2nd}
\end{align}
Note that $W$ coincides with the Wronskian defined for the homogenous solutions.
Therefore, $G$ in Eq.~\eqref{Green} is the Green's function constructed from the Wronskian method when we solve the inhomogeneous classical equation of motion, and $S$ in Eq.~\eqref{source2nd} corresponds to the second order source of the scalar field fluctuations in the classical equation of motion.
Therefore, we may regard Eq.~\eqref{source2nd} as the induced GWs.
Let us consider the IR behavior of the induced GWs.
For $q\ll p_2,p_3$ and $|q\tau|,|q\tau'|\ll 1$, we find 
\begin{align}
	S^s_{\mathbf q}[\delta \chi]= 	e^{ij,s,*}(\hat q)\Sigma_{ij}[\Xi],
\end{align}
where $\Sigma_{ij}$ is a $\mathbf q$ independent functional of $\Xi$, and the Green's function is expanded as
\begin{align}
	G_q(0;\tau') = -\frac{4 H^2 a'^2\tau'^3}{3M_{\rm pl}^22}\left[ 1 +\mathcal O(\tau'^2q^2)\right].\label{Gind}
\end{align}
Let us define the power spectrum of the induced GW as
\begin{align}
	\left. \langle 0| \sum_{s} h^s_{\mathbf q,{\rm ind.}}(\tau) h^s_{\bar{\mathbf q},{\rm ind.}}(\tau)  |0 \rangle \right|_{\tau=0} \equiv (2\pi)^3\delta(\mathbf q+\bar {\mathbf q})P_{\rm ind}.
\end{align}
The leading order $q$ independence of $G$ and $S$ implies
\begin{align}
	\frac{P_{\rm ind}}{P_{h0}} =\mathcal O(\tilde q^3), 	
\end{align}
which is the well-known IR behavior of the causally generated induced tensor spectrum.
Thus, $P_{\rm ind.}=\mathcal O(\lambda^2)$ so it is included in $P_{h2}$, but the induced GWs do not contribute to the super-horizon enhancement of the one-loop tensor spectrum as they are causally produced.
An explicit form of the auto-power spectrum of the induced tensor mode at $\tau=0$ is given as
\begin{align}
\frac{P_{\rm ind}}{P_{h0}}=	&-\frac{1}{4}  \int^0_{\tau_i} d\tau' a(\tau')^2 \int^{\tau'}_{\tau_i} d\tau''
	 a(\tau'')^2 
	 \notag \\
	 &\times
	 (v_{q}(\tau')-v^*_{q}(\tau'))(v_{q}(\tau'')-v^*_{q}(\tau''))
	 \notag \\
	 &\times 
	 \int_0^{\infty} d p_2 \int_{|p_2-q|}^{p_2+q} dp_3  \bar w(q; p_2,p_3) 
	\notag \\
	&\times  
	\left( u_{p_2}(\tau')u^*_{p_2}(\tau'')u_{p_3}(\tau')u^*_{p_3}(\tau'')
	\right.
	\notag \\
	&\left.+
	u_{p_2}(\tau'')u^*_{p_2}(\tau')u_{p_3}(\tau'')u^*_{p_3}(\tau')
	\right).
\end{align}
Note that $P_{\rm ind}$ is strictly non-negative as it is a squared quantity.
Then, for the delta-function-like scalar source, we find
\begin{align}
&\frac{P^\delta_{\rm ind}}{P_{h0}}=	- \frac{p_*^4(4p_*-q^2)^2}{512\pi^2 q}\Theta_{2-q/p_*}  
\notag \\
&\times \int^0_{\tau_0} d\tau' a(\tau')^2 \int^{\tau'}_{\tau_0} d\tau''
	 a(\tau'')^2 
	\notag \\
	&\times  
	(v_{q}(\tau')-v^*_{q}(\tau')) (v_{q}(\tau'')-v^*_{q}(\tau'')) 
	\notag \\
	&\times (
	 u^2_{p_*}(\tau')(u^*_{p_*}(\tau''))^2
	 +
	 u^2_{p_*}(\tau'')(u^*_{p_*}(\tau'))^2
	 ),
\end{align}
where the lognormal case is given by $\Theta \to \Theta^{\Delta}$. 
The induced tensor modes are presented in Figs.~\ref{pow} and \ref{cut_fig}.

Another $\mathcal O(\lambda^1)$ correction originates from the fourth-order interaction Hamiltonian:
\begin{align}
&	i \int^\tau_{\tau_i} d\tau'	[H^{(4)}_{{\rm int}, I}(\tau') ,h^s_{\mathbf q}(\tau)]
		\notag \\
		&=
	 \frac{i}{2}\int^\tau_{\tau_i} d\tau' a(\tau')^2[v_{q}(\tau)v^*_{q}(\tau')- v_{q}(\tau')v^*_{q}(\tau)]\notag \\
	 &\times 
	 \prod_{i=2}^4\left(\int \frac{d^3p_i}{(2\pi)^{3}}\right)\delta\left(\sum_{i=2}^4 \mathbf p_i-\mathbf q\right) \sum_{s_2}  
\notag 	 \\
	 &\times   e^{s*}_{ik}(\hat q)e^{s_2}_{kj}(\hat p_2) p_{3i}p_{4j}   \delta \chi_{\mathbf p_3}(\tau')  \delta \chi_{\mathbf p_4}(\tau')
	h^{s_2}_{\mathbf p_2}(\tau')
	.
\end{align}
Taking the partial average over the scalar field, we can extract the component parallel to $h^s_{\mathbf q}$ as
\begin{align}
&	i \int^\tau_{\tau_i} d\tau'	\langle [H^{(4)}_{{\rm int}, I}(\tau'),h^s_{\mathbf q}(\tau)]\rangle_{\delta \chi}
	\notag \\
			=&
	\frac{i}{6}\int^\tau_{\tau_i} d\tau' a(\tau')^2( v_{q}(\tau)v^*_{q}(\tau') - v^*_{q}(\tau)v_{q}(\tau') )
	 \notag 
	 \\
	 &\times \int \frac{p_3^4dp_3}{2\pi^{2}} |u_{p_3}(\tau')|^2  h^{s}_{\mathbf q}(\tau')
	,\label{H4tens}
\end{align}
where the subscript $\delta \chi$ implies that we integrated out $\delta \chi$.
The cross-correlation of Eq.~\eqref{H4tens} and $h^s_{\mathbf q}$ reproduces the result in the in-in formalism~\eqref{P13}.

The variation of the super-horizon tensor fluctuation can be understood in a quantum language as follows.
Expanding Eq.~\eqref{H4tens} in terms of the tensor annihilation and creation operators, we find
\begin{align}
	&h^s_{\mathbf q}(\tau)+i \int^\tau_{\tau_i} d\tau'	\langle [H^{(4)}_{{\rm int}, I}(\tau') ,h^s_{\mathbf q}(\tau)]\rangle_{\delta \chi} \notag \\
	=&V_q(\tau) b^{s}_{\mathbf q} + V_q^*(\tau) b^{s\dagger}_{-\mathbf q},\label{H4super}
\end{align}
where we introduced the new positive frequency mode function
\begin{align}
		V_q(\tau) = \alpha(\tau) v_{q}(\tau) + \beta(\tau) v^*_{q}(\tau),\label{lintransfforV}
\end{align}
with
\begin{align}
	\alpha_q(\tau) &  \equiv 1 +  \frac{i}{6}\int^\tau_{\tau_i} d\tau' a(\tau')^2 \int \frac{p_3^4dp_3}{2\pi^{2}} |u_{p_3}(\tau')|^2  |v_q(\tau')|^2,
	\\
	\beta_q(\tau) & \equiv -\frac{i}{6}\int^\tau_{\tau_i} d\tau' a(\tau')^2\int \frac{p_3^4dp_3}{2\pi^{2}} |u_{p_3}(\tau')|^2 v^2_{q}(\tau').
\end{align}
Then $\alpha$ and $\beta$ satisfy the following relation up to one-loop order
\begin{align}
	|\alpha_q |^2 - |\beta_q|^2 =1. \label{4jicoef}
\end{align}
Therefore, the linear transformation from $(v,v^*)$ to $(V,V^*)$ in Eq.~\eqref{lintransfforV} can be understood as a Bogoliubov transformation at least up to one-loop order.
The power spectrum is written as
\begin{align}
	\frac{P_{h1}}{P_{h0}} = 2|\beta_q(0)|.
\end{align}

\medskip
A Bogoliubov transformation also appears in $\mathcal O(\lambda^2)$ correction.
$\mathcal O(\lambda^2)$ terms include mode mixing of tensor fluctuations, but we can extract the component proportional to $h^s_{\mathbf q}$ by taking the average over $\delta \chi$.
After some algebra, we find
\begin{align}
	&h^s_{\mathbf q}(\tau) -\int^\tau_{\tau_i} d\tau' \int^{\tau'}_{\tau_i} d\tau'' \langle [H^{(3)}_{{\rm int},I}(\tau''),[H^{(3)}_{{\rm int},I}(\tau'),h^s_{\mathbf q}(\tau)]] \rangle_{\delta \chi}
	\notag \\
	&=\tilde V_q(\tau) b^s_{\mathbf q} + \tilde V_q^*(\tau)  b^{s\dagger}_{-\mathbf q} \label{119},
\end{align}
where defined
\begin{align}
	\tilde V_q (\tau)= \tilde \alpha_q(\tau) v_q(\tau) + \tilde \beta_q(\tau) v^*_q(\tau),\label{transfbar}
\end{align}
with 
\begin{align}
		\tilde \alpha_q &=  1 + \tilde \beta_q^*,
		\label{alphatilde}
		\\
		\tilde \beta_q &=\frac14 \int^\tau_{\tau_0} d\tau' a(\tau')^2 \int^{\tau'}_{\tau_0} d\tau''  a(\tau'')^2  
		\notag 
		\\
	&\times   \int^\infty_0dp_2 \int^{p_2+q}_{|p_2-q|} dp_3 \bar w(q; p_2,p_3) v_q^*(\tau'')v_{q}(\tau')
	 \notag 
	 \\
	 &\times  (u^*_{p_3}(\tau'') u_{p_3}(\tau')- u_{p_3}(\tau'') u^*_{p_3}(\tau'))
	 \notag \\
	 &\times (u_{p_2}(\tau'')u^*_{p_2}(\tau')+u_{p_2}(\tau')u^*_{p_2}(\tau'')).
		\label{betatilde}
\end{align}
Then, we find
\begin{align}
	&|\tilde \alpha_q|^2 - |\tilde \beta_q|^2 = 1+ \tilde \beta_q+\tilde \beta_q^*.\label{123}
\end{align}
The one-loop term in Eq.~\eqref{123} does not straightforwardly vanish since, in general, $v_q(\tau'')\neq v_q(\tau')$ unlike Eq.~\eqref{4jicoef}.
For the IR mode $|q\tau| \ll1$, when $|q\tau'|,|q\tau''|\ll 1$, the time dependence of the tensor mode functions is dropped and we get $v_q(\tau'')\sim v_q(\tau')$.
In this case, we can also think of the linear transformation~\eqref{transfbar} as the Bogoliubov transformation up to one-loop order.

We concretely showed we have Bogoliubov transformations for several limited circumstances, but Eq.~\eqref{Heisen-hop} straightforwardly implies
\begin{align}
	h^s_{\mathbf q,{\rm H}}(\tau) &= v_q(\tau) B^{s}_{\mathbf q}(\tau) + v^*_q(\tau) B^{s,\dagger}_{-\mathbf q}(\tau),
	\\
	B^s_{\mathbf q}(\tau) &\equiv  F^{-1}(\tau) b^s_{\mathbf q} F(\tau),
\end{align}
The unitarity of $F$ leads to
\begin{align}
	[B^{s}_{\mathbf q},B^{\bar s}_{-\bar{\mathbf q}}] = [F^{-1} b^s_{\mathbf q} F,F^{-1} b^{s\dagger }_{-\mathbf q} F]=[b^s_{\mathbf q} , b^{s\dagger }_{-\mathbf q}]\label{btoB}.
\end{align}
Thus, the transformation $b\to B$ is a generalized Bogoliubov transformation.
The partial trace in terms of $\delta \chi$ might break the unitarity in general, i.e., the Eq.~\eqref{btoB} is not necessarily established once we take the average of $\delta \chi$. At the same time, we confirmed that the unitarity is kept at the one-loop order for the IR limit. 
$F^{-1} b^s_{\mathbf q} F$ is not necessarily written by $b_{\mathbf q}$ and $b^\dagger_{\mathbf q}$ only as $F$ contains the mode coupling terms between different Fourier modes and scalar fluctuations.
Generally, the present Bogoliubov transformation mixes the Fourier modes and $\delta \chi$, and the Bogoliubov coefficients are operator-valued in the presence of $\delta\chi$, which satisfies the generalized conditions for the multi-variable Bogoliubov transformation.
We leave the investigation for future work.

The Bogoliubov perspective also explains why the super-horizon scalar fluctuations do not contribute to the loop correction, e.g., (D) in Fig.~\ref{pow}.
For super-horizon scalar fields, i.e., when  $p_*\tau',p_*\tau''\ll 1$, we have $u_{p_*}=u_{p_*}^*$.
Then we find $\tilde \alpha_q=1$ and $\tilde \beta_q=0$, which is why the super-horizon enhancement of the spectator scalar field does not contribute to the loop effect.
Ref.~\cite{Kanno:2018cuk} claimed that interactions with external classical fields generate a coherent state as the unitary operator is written in a form of the displacement operator.
However, the claim does not apply to the present case since the interacting scalar field is neither external nor classical.

\subsection{Bogoliubov transformation for scalar amplitude enhancement}
So far, we have discussed that the loop correction can be understood as the Bogoliubov transformation for tensor fluctuations.
A Bogoliubov transformation may also model a class of enhanced spectator scalar fields for $\delta \chi$ rather than a simple multiplicative enhancement of the scalar field~\eqref{expamp}.
Enhancement of $\delta \chi$ via a unitary evolution is categorized into this class~\footnote{
Recent work~\cite{Inomata:2022yte} considered inflationary scalar one-loop calculation for the oscillatory features in the inflaton potential. They numerically obtained the mode function by directly integrating the linear equation of motion. Then the enhancement feature looks similar to the enhancement via the Bogoliubov transformation.
}.
In this case, the excited state of the spectator field can be written as
\begin{align}
	\delta \chi_{\mathbf q} = U_q a_{\mathbf q} + U^*_q a_{-\mathbf q}^\dagger,\label{BTchi}
\end{align}
where we consider
\begin{align}
	U_q = \frac{e^{i\phi_1}}{2}\left(\Xi+\frac{1}{\Xi}\right) u_q+ \frac{e^{i\phi_2}}{2}\left(\Xi-\frac{1}{\Xi}\right) u_q^*.\label{single:BT}
\end{align}
With the physical time we have
\begin{align}
	\frac{\Xi+\Xi^{-1}}{2} &= \cosh[ \mu H (t-t_i) ],~
	\\
	\frac{\Xi-\Xi^{-1}}{2} &= \sinh[ \mu H (t-t_i) ].
\end{align}
In this parameterization, the power spectrum of the spectator scalar field is amplified by $\Xi^2$, i.e., $|U_q|^2/|u_q|^2 = \Xi^2$ for $\phi_1=\phi_2$. 
Then, all equations above can be reused after $u_q\to U_q$.
Note that Eq.~\eqref{single:BT} is a most general parameterization of the Bogoliubov transformation for a single field.
$P_{h1}$ does not change much for the new parameterization, while $P_{h2}$ changes drastically.
Large $\Xi$ leads to $U_q\sim U_q^*$.
We should note that the first iterative component disappears, i.e., $\tilde \alpha_q= 1 $ and $\tilde \beta_q=0$ in Eqs.~\eqref{alphatilde} and \eqref{betatilde} if $U_q= U_q^*$. 
Therefore, the violation of $U_q= U_q^*$ introduces non-vanishing scale invariant enhancement of the loop.
Hence, in Eq.~\eqref{single:BT}, the appearance of $\Xi^{-1}$ terms is crucial.
In the new parameterization, we analytically find $\Gamma$ in Eq.~\eqref{deltaPint}
is replaced as 
\begin{align}
&\tilde \Gamma[\Xi] = 2\int^{0}_{x_i} dx' \int^{x'}_{x_i} dx''\frac{x'}{x''^2}  
\notag \\
&\times (x' \sin x'+\cos x') (x'' \sin x''+\cos x'') 
\notag \\
&\times \left[\Xi''^2  (x' \cos x'-\sin x') (x'' \sin x''+\cos x'')\right.
\notag 
\\
&\left.-\Xi'^2  (x' \sin x'+\cos x') (x'' \cos x''-\sin x'')\right].
\label{deltaPint_BT}
\end{align}
Thus, the leading order contribution $\mathcal O(\Xi^4)$ of the first iterative correction is canceled, and the enhancement factor becomes $\mathcal O(\Xi^2)$.
This cancellation is also seen from Eq.~\eqref{betatilde} as we have
\begin{align}
	&U^*_{p_3}(\tau'') U_{p_3}(\tau')- U_{p_3}(\tau'') U^*_{p_3}(\tau') 
	\notag \\
	=& u^*_{p_3}(\tau'') u_{p_3}(\tau')- u_{p_3}(\tau'') u^*_{p_3}(\tau').
\end{align}

\begin{figure*}
  \includegraphics[width=\linewidth]{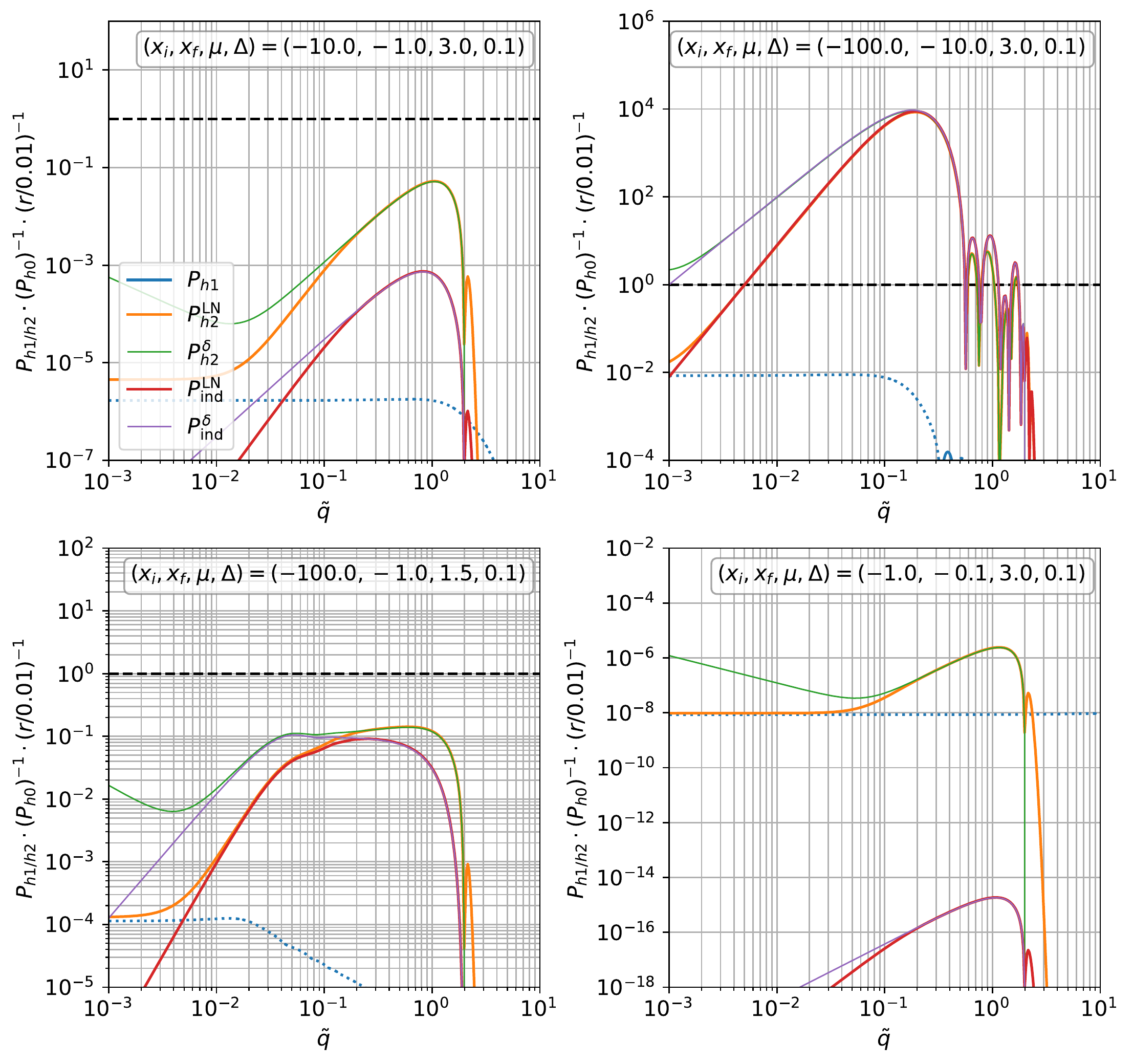}
  \caption{One-loop spectrum for the scalar field excited via the Bogoliubov transformation~\eqref{BTchi}. Definitions of symbols are the same as Fig.~\ref{pow}. The magnitude of $P_{h1}$ and $P_{h2}$ are comparable as suggested by Eq.~\eqref{deltaPint_BT}.}
  \label{BT_fig}	
\end{figure*}

In the present setup, the magnitude of $P_{h1}$ and $P_{h2}$ are comparable in the IR region, and their signs can be opposite. The exact cancellation should not happen as we compare the different interactions.
Numerical calculation for (A) to (D) in the new parameterization is presented in Fig.~\ref{BT_fig}.
The super-horizon enhancement is relatively suppressed, as we discussed.
(B) implies the secondary spectrum is dominated by the induced tensor modes for subhorizon scalar fields, while (A) and (D) suggest near horizon scalar fluctuation hardly produces the induced tensor fluctuations in contrast to Figs.~\ref{pow} and \ref{cut_fig}.

\medskip
For multi-field inflation, Bogoliubov transformation for a scalar field is not necessarily written as Eq.~\eqref{single:BT}.
Eq.~\eqref{BTchi} contains different operators, and we cannot use the above results straightforwardly.
We leave the detailed analysis for the multi-spectator fields for future work.

\section{Separate universe approach in Bianchi type-I cosmology}
\label{bianchi1}

We considered several quantum loops during inflation, using the in-in formalism.
The new contributions change the super-horizon tensor spectrum, which may sound odd from causality.
We claimed that the time evolution could be regarded as a Bogoliubov transformation.
It just changes the definition of the vacuum state via the nonlinear interaction so that the super-horizon tensor fluctuations are seen differently without violation of causality.

For skeptical readers, we provide another look at our results in this section, focusing on the classical counterpart of the effects.
We may understand the classical evolution of super-horizon perturbations from the separate Universe approach.
We can think of different realizations of local FLRW universes as fluctuations of e-folding numbers, i.e., the super-horizon curvature perturbations. 
Similarly, realizations of super-horizon tensor fluctuations may be regarded as those of different homogenous but anisotropic universes, i.e., Bianchi type-I spacetimes, whose metric may be written as
\begin{align}
	ds^2 = a^2\left(-d\tau^2  + e^{2b} dx^2 +e^{-2b} dy^2 + dz^2 \right).
	\label{dsBianchi}
\end{align}
When expanding Eq.~\eqref{dsBianchi} to linear order in $b$, one finds that $b$ coincides with the realization of a plus mode propagating along the $z$ axis.
The transverse condition holds non-perturbatively as the components are in the plane perpendicular to the propagating direction.
On the other hand, the traceless condition is not well-defined without a background spacetime metric~\cite{Ota:2021fdv}.
The physical implication of the traceless condition is that the volume element is not affected by the perturbation.
In this sense, $b$ is properly defined such that the volume element is not perturbed to all orders in Eq.~\eqref{dsBianchi}.
Hence we consider Eq.~\eqref{dsBianchi} as a nonperturbative extension of the tensor perturbation in the separate Universe approach.
Eq.~\eqref{dsBianchi} can also be obtained by focusing on a single mode of $h_{ij}$ in Eq.~\eqref{gijdef} and diagonalizing it.
In the anisotropic spacetime, we consider a minimally coupled classical scalar field $\chi$ whose action is written as
\begin{align}
	S_\chi =-\frac{1}{2} \int d^4 x \sqrt{-g}g^{\mu\nu}\partial_\mu \chi \partial_\nu \chi+\cdots,
\end{align}
where $\cdots$ implies the model-dependent potential terms, which determine the details of the scalar dynamics.

Let us consider the effect of the scalar field on the spacetime anisotropy $b$.
First, the Einstein tensor is written as~(without perturbative expansion for $b$)
\begin{align}
	G_{00}  &=  3\mathcal H^2   -  \dot b^2, \\
	G_{11}  &=  e^{2b}(-\mathcal H^2  -2 \dot{\mathcal H} + 2\mathcal H\dot b + \ddot b - \dot b^2) ,\\
	G_{22}  &=  e^{-2b}(-\mathcal H^2  -2 \dot{\mathcal H} - 2\mathcal H \dot b - \ddot b  - \dot b^2), \\
	G_{33}  &=  -\mathcal H^2  -2 \dot{\mathcal H}  - \dot b^2,
\end{align}
where the overdot is the derivative with respect to the conformal time, and we introduced the conformal Hubble parameter $\mathcal H\equiv \dot a/a$, and then the energy-momentum tensor is given as
\begin{align}
	T_{00}&= \partial_0\chi \partial_0\chi + \frac{1}{2} a^{2}(X+\cdots)   ,\\
	T_{11}&= \partial_1\chi \partial_1\chi -\frac{1}{2} a^{2}e^{2b}(X+\cdots)  ,\\
	T_{22}&=  \partial_2\chi \partial_2\chi-\frac{1}{2} a^{2}e^{-2b}(X+\cdots)  ,\\
	T_{33}&= \partial_3\chi \partial_3\chi -\frac{1}{2} a^{2}(X+\cdots)  ,
\end{align}
and we defined 
\begin{align}
	X\equiv g^{\rho\sigma}\partial_\rho \chi \partial_\sigma \chi.
\end{align}
Then, the Einstein equation yields
\begin{align}
	\ddot b +2 \mathcal H\dot b = \frac{e^{-2b}\partial_1\chi \partial_1\chi - e^{2b}\partial_2\chi \partial_2\chi }{2 M_{\rm pl}^2}.\label{beomderived}
\end{align}
Thus, the model-dependent terms do not contribute to the equation of motion for $b$.
We limit our calculation for the Born approximation for simplicity; we only consider the classical counterpart of $P_{h1}$-like contribution.
When expanding $\chi$ with iterative solutions, additional terms appear in Eq.~\eqref{beomderived} for consistent iteration.

Without the scalar field, we get
\begin{align}
	b =  c_1 \int^\tau \frac{d\tau'}{ a(\tau')^2} + c_2.
\end{align}
The constant solution $c_2$ corresponds to the super-horizon conserved tensor mode. 
$c_1$ is the decaying mode. 

For a non-vanishing scalar field, statistical isotropy of $\chi$ implies that we have
\begin{align}
	\langle \partial_x \chi \partial_x  \chi \rangle =\langle \partial_y  \chi \partial_y  \chi \rangle =\frac{1}{3} \langle (\partial_i \chi)^2 \rangle \geq 0.
\end{align}
the leading order EoM for $b$ becomes
\begin{align}
	 b'' +2 \mathcal Hb' +m_{\rm eff}^2b  = \mathcal O(b^2),\label{meffeom}
\end{align}
where we defined
\begin{align}
	m_{\rm eff}^2 = \frac{2\langle (\partial_i \chi)^2 \rangle}{3M_{\rm pl}^2}\geq 0.
\end{align}
Thus, the back reaction of $\chi$ behaves like a mass term of the tensor mode $b$.
This is a harmonic oscillator in a friction force with an effective mass $m_{\rm eff}$, which results in the decaying solution whose time scale is given by $m_{\rm eff}^{-1}$.
Suppose we have a scalar field enhancement at $p=p_*$ mode during inflation. We have
\begin{align}
	m_{\rm eff}^2 = \frac{1}{3}\frac{H^2}{M_{\rm pl}^2} (1 + \tau^2 p_*^2)p_*^2\Xi^2. 
\end{align}
The effective mass squared is suppressed by a factor of $H^2/M_{\rm pl}^2$, so the mass is usually negligibly small.
$b$ is a realization of super-horizon tensor fluctuations that exit the horizon in the early stage of inflation, which is constant at the initial time.
Once we get a large $\Xi$ when the enhanced scalar mode is inside the horizon $|p_*\tau|\gtrsim 1$, we get $|m_{\rm eff} \tau|\gtrsim 1$, and $b$ oscillates and decays.
Thus, we showed that the background anisotropy could evolve due to the scalar field.
Then the super-horizon suppression of $P_{h1}$ is explained in the separate Universe approach.
A similar argument will be possible for $P_{h2}$, including the first iterative solution in $\chi$.
Solving Eq.~\eqref{meffeom} is a non-perturbative way to find the super-horizon variation.
For example, we can find the solution for constant $m_{\rm eff}$, which is given by a linear combination of the Bessel functions with the time constant $m_{\rm eff}$.
The Bessel functions account for the effects of $m^2_{\rm eff}\propto  H^2/M_{\rm pl}^2$ to all order non perturbatively.

\section{Observations}
\label{obs:sec}

One normally considers that scalar fields during inflation decay in the end to realize the radiation-dominant Universe.
Hence $\delta \chi$ should decay at some point, and the one-loop correction to the tensor fluctuation is fixed.
The decay of $\delta \chi$ is model-dependent, which is not considered in this paper.
Once $\delta \chi$ decays, the time evolution of the tensor fluctuations afterward should be the same as the linear tensor modes.
The tensor fluctuations with the loop re-enter the horizon and propagate as gravitational waves.
If large curvature perturbations are produced by the end of inflation from the large $\delta \chi$, a similar GW reduction or enhancement will happen in the Universe after reheating~\cite{Chen:2022dah}.

We can apply the standard CMB polarization analysis for the corrected gravitational waves since the time evolution of the GWs is the same as the linear one in the late universe.
However, the tensor-to-scalar-ratio constrained in the observation should be related to
\begin{align}
	r_{\rm NL} = \frac{P_{h0}+ P_{h1} + P_{h2}+\cdots}{P_{\zeta0}+ P_{\zeta1} + P_{\zeta2}+\cdots},
\end{align}
where we also expect the loop correction for the adiabatic perturbations for the same reason.
The loop correction for $\zeta$ is model-dependent.
Also, the coupling to $\zeta$ is more slow-roll suppressed, so that the correction may be relatively small.

Calculation of $\Omega_{\rm GW} \equiv \rho_{\rm GW}/\rho_{\rm tot}$ with the GW energy density $\rho_{\rm GW}$ and the total energy density $\rho_{\rm tot}$ can be done as if they are linear GWs; the energy density of GWs are computed as~\cite{Isaacson:1968zza,Isaacson:1968hbi} 
\begin{align}
	\frac{d\ln \rho_{\rm GW}}{d\ln q}& =\frac{M_{\rm pl}^2}{4a^2}  \left(\frac{dT_{h}(q\tau)}{d \tau}\right)^2 \frac{q^3(P_{h0} + P_{h1} + P_{h2} +\cdots )}{2\pi^2},
\end{align}
where $T_h$ is the linear transfer function of the tensor mode.
Given an enhancement model of the scalar amplitude, we can put some constraints on the parameters from the current or future measurements of $r_{\rm NL}$ and $\Omega_{\rm GW}$.

\medskip
A remaining issue is the gauge dependence of the loop contribution.
Tensor fluctuations are gauge-independent at linear order but are not at nonlinear order.
The gauge issue has been discussed in the context of induced GWs in the literature~\cite{Inomata:2019yww,DeLuca:2019ufz,Domenech:2020xin,Ota:2021fdv}.
Only the propagating component of the tensor modes can be regarded as GWs, which are gauge independent in the subhorizon limit, as far as we choose a gauge condition where the metric perturbations do not diverge in the short scale.
We expect similar arguments for the nonlinear interaction.
We showed that the super-horizon scalar field hardly contributes to the loop correction.
Therefore, as far as the gauge-dependent component of $\delta \chi$ is small before horizon exit, we expect the gauge independent loop.
In the in-in formalism, equations are expanded by the linear perturbations.
Therefore, gauge transformation can be truncated at linear order.
We can always construct the linear gauge invariant from the gauge-dependent perturbations.
The gauge invariants can express our Hamiltonian, and the loop correction should be gauge invariant.

\section{Conclusions}
\label{conc}

This paper presented a full-quantum calculation of the one-loop inflationary tensor power spectrum in the presence of an excited spectator scalar field.
We found that the one-loop correction from an excited subhorizon scalar mode may enhance or reduce the super-horizon primordial tensor power spectrum scale-invariantly.
The first impression of the results is odd, as we believe that the super-horizon tensor fluctuations should be constant until horizon re-entry from causality.
Our setup differs from the previous literature about induced GWs based on the classical equation of motion.
We employed the in-in formalism, and the full one-loop effect was considered.
Our calculation includes the causal production of GWs, limited to the subhorizon scale, as we expected.
The super horizon loop effects are considered as Bogoliubov transformation.
Nonlinear interaction among the tensor and scalar changes the definition of the vacuum, and then the initial super horizon tensor fluctuations are observed differently.
We examined the separate Universe approach with Bianchi type-I local universes to explain the super-horizon evolution of tensor fluctuations.
We find that the background anisotropy reduces when a scalar field exists, a damping effect introduced by an effective mass.
This is the separate Universe counterpart of $P_{h1}$.

We considered several forms of enhanced scalar fields.
For example, a simple multiplication of the mode functions by $\Xi$ leads to the loop correction of $\mathcal O(\Xi^4)$.
This amplification may happen when, e.g., we modify the sound speed.
 In contrast, we found a leading order cancellation for a scalar enhancement via the Bogoliubov transformation, which results in $\mathcal O(\Xi^2)$ in the super Hubble region.
The size of the one-loop correction is sensitive to the amplitude and the initial and final amplification time. 
We saw several physically different situations (A) to (D) in section~\ref{model:sec}; depending on the situations, the final spectrum varies.
Hence, we need to specify a model and parameters for quantitative analysis.
We found that the loop correction may exceed the tree level power spectrum, which indicates the breakdown of the perturbative expansion in the presence of a highly enhanced scalar field.

Once the scalar peak disappears, the super-horizon correction is fixed, and we cannot distinguish it from the original primordial GWs.
Time evolution after the horizon re-entry is linear, and observational constraints on the linear tensor power spectrum straightforwardly apply to the corrected spectrum.

Before concluding this paper, let us discuss some interesting extensions of this work.
First, our calculation is based on several toy models of an enhancement, and the relationship between $\delta \chi$ and the final $\zeta$ is unknown. 
Considering the loop effect for a concrete model with scalar amplitude enhancement should be interesting since we can predict the relationship between the PBH formation history and the one-loop inflationary spectrum~(see, e.g. recent works~\cite{Kristiano:2022maq, Inomata:2022yte}.). 
Computing a one-loop bispectrum or trispectrum should also be interesting as the loop correction suggests large non-Gaussianity for the tensor fluctuations.
This property could be useful to distinguish the linear tensor power spectrum from the one-loop.
As we mentioned in Section~\ref{sec2:w}, the (in)equivalence of our quantum approach and classical field theory with the stochastic initial condition is not obvious.
As far as we compute the induced GWs in the quantum approach, the spectrum looks quite similar.
Solving a consistent iteration in the EoM may tell us the quantum nature during inflation.
The loop correction for the scalar power spectrum should be technically complicated since it is fully model-dependent. However, we expect the same momentum structure for the one-loop terms for the scalar power spectrum.
Our result suggests that large-scale measurements may indirectly test the short-scale enhancement of cosmological perturbations, so combining gravitational wave detectors at all scales is crucial for future surveys~\cite{Kogut:2011xw,Andre:2013afa,Matsumura:2013aja,Desvignes:2016yex,Brazier:2019mmu,Kerr:2020qdo,LISA:2017pwj,Kawamura:2011zz,Ruan:2018tsw,TianQin:2015yph}.

\begin{acknowledgments}
The authors would like to thank Keisuke Inomata for useful comments.
AO would like to thank Chen Chao, Yuhang Zhu and Huiyu Zhu for useful discussions.
AO and YW are supported in part by the National Key R\&D Program of China (2021YFC2203100), the NSFC Excellent Young Scientist Scheme (Hong Kong and Macau) Grant No.~12022516, and the RGC of Hong Kong SAR, China (GRF 16303621).
MS is partly supported by the JSPS KAKENHI grant Nos.~19H01895, 20H04727, and 20H05853.

\end{acknowledgments}

\appendix

\section{Operator products}
In this paper, we often use the following products of the scalar and tensor fluctuations:
\begin{align}
	\langle 0| \delta \chi_{\mathbf p_1}  \delta \chi_{\mathbf p_2}  	|0\rangle
	&=(2\pi)^3\delta(\mathbf p_1+\mathbf p_2)
	 u_{p_1}u^*_{p_2} 
	 \label{58},
	 \\
	 	\langle 0| h^{s_1}_{\mathbf p_1}  h^{s_2}_{\mathbf p_2}  	|0\rangle
	&=(2\pi)^3\delta(\mathbf p_1+\mathbf p_2)\delta^{s_1s_2}
	 v_{p_1}v^*_{p_2}.\label{58.5} 
\end{align}
Also, the products of the four operators are given as
\begin{align}
	&\langle 0| \delta \chi_{\mathbf p_1}  \delta \chi_{\mathbf p_2}   \delta \chi_{\mathbf p_3}  \delta \chi_{\mathbf p_4}	|0\rangle
	\notag \\
	=&(2\pi)^6\delta(\mathbf p_1+\mathbf p_2)
	\delta(\mathbf p_3+\mathbf p_4) u_{p_1}u^*_{p_2}u_{p_3}u^*_{p_4} 
	\notag \\
	+&(2\pi)^6\delta(\mathbf p_1+\mathbf p_3)
	\delta(\mathbf p_2+\mathbf p_4)u_{p_1}u^*_{p_3}u_{p_2}u^*_{p_4}
	\notag \\
	+&(2\pi)^6\delta(\mathbf p_1+\mathbf p_4)
	\delta(\mathbf p_2+\mathbf p_3)u_{p_1}u^*_{p_4}u_{p_2}u^*_{p_3} \label{59},
\end{align}
and
\begin{align}
	&\langle 0| h^{s_1}_{\mathbf p_1}h^{s_2}_{\mathbf p_2}h^{s_3}_{\mathbf p_3} h^{s_4}_{\mathbf p_4}|0\rangle \notag \\
	=&(2\pi)^6\delta(\mathbf p_1+\mathbf p_2)
	\delta(\mathbf p_3+\mathbf p_4) \delta^{s_1s_2}\delta^{s_3s_4}v_{p_1}v^*_{p_2}v_{p_3}v^*_{p_4} 
	\notag \\
	+&(2\pi)^6\delta(\mathbf p_1+\mathbf p_3)
	\delta(\mathbf p_2+\mathbf p_4)\delta^{s_1s_3}\delta^{s_2s_4} v_{p_1}v^*_{p_3}v_{p_2}v^*_{p_4}
	\notag \\
	+&(2\pi)^6\delta(\mathbf p_1+\mathbf p_4)
	\delta(\mathbf p_2+\mathbf p_3)\delta^{s_1s_4}\delta^{s_2s_3} v_{p_1}v^*_{p_4}v_{p_2}v^*_{p_3} .\label{60}
\end{align}

\section{Angular integrals}\label{angular_int}
In this appendix, we discuss the treatment for the polarization sum and the angular dependence in the loop integrals.
Eq.~\eqref{64} can be computed in a convenient coordinate system where $\hat q=(0,0,1)$.
In this frame, the polarization tensor can be written as (up to a phase factor)
\begin{align}
	e_{ij}^{\pm2}(\hat q) \equiv \frac{1}{2}\begin{pmatrix}
 1 & \pm i & 0 \\
 \pm i & -1 & 0  \\
 0  & 0 & 0
\end{pmatrix},
\end{align}
and then the rest of the momentum vectors are 
\begin{align}
	\mathbf p_A = 	 p_A \begin{pmatrix}
 \sin \theta_A \cos \phi_A & \sin \theta_A \sin \phi_A & \cos\theta_A
\end{pmatrix}
\end{align}
Using these specific representations, we find
\begin{align}
	 \sum_{s=\pm2} |e^{ij,s}(\hat q) p_{2i}p_{3j} |^2 
	 &
=\frac{p_2^2p_3^2}{2}\sin^2\theta_2\sin^2\theta_3,\label{sinsin}
\end{align}
which can be written by $p_2$ and $p_3$ since we have
\begin{align}
	\sin^2\theta_2\sin^2\theta_3 &= \frac{1}{16p_2^2p_3^2q^4} (p_2+p_3+q)^2(p_2+p_3-q)^2
	\notag \\
	&\times (p_2-p_3+q)^2(-p_2+p_3+q)^2.\label{sinsin2}
\end{align}
Thus, momentum integrals in Eq.~\eqref{64} 
is only a function of $p_2$ and $p_3$.
When the integrand is independent of the angular coordinates, we have
\begin{align}
				&\int \frac{d^3p_2d^3p_3}{(2\pi)^6}(2\pi)^3\delta(\mathbf q-\mathbf p_2-\mathbf p_3) 
				\notag \\
=&		\frac{1}{(2\pi)^2q}\int_0^{\infty} d p_2 \int_{|p_2-q|}^{p_2+q} dp_3 p_2p_3 .\label{wakeru}
\end{align}
Combining Eqs.~\eqref{sinsin}, \eqref{sinsin2}, and \eqref{wakeru}, we integrate out the angular coordinates.
A derivation of Eq.~\eqref{wakeru} is given as
\begin{align}
		&\int \frac{d^3p_2d^3p_3}{(2\pi)^6}(2\pi)^3\delta(\mathbf q-\mathbf p_2-\mathbf p_3) 
	\notag 	 \\
	=&\int d^3 x  e^{i\mathbf q\cdot \mathbf x} \int \frac{d^3p_2d^3p_3}{(2\pi)^6}e^{i\mathbf p_2\cdot \mathbf x}e^{i\mathbf p_3\cdot \mathbf x}  
	\notag	\\
	=&(4\pi)^2\sum_{\ell \ell_2\ell_3 m m_2m_3}i^{\ell_2+\ell_3}\int d^3 x \int \frac{d^3p_2d^3p_3}{(2\pi)^6} 
	\notag \\
	&\times  Y_{\ell m}(\hat x)Y_{\ell_2 m_2}(\hat x)Y_{\ell_3 m_3}(\hat x)
	 Y_{\ell m}^\star (\hat q)   Y_{\ell_2 m_2}(\hat p_2) Y_{\ell_3 m_3}(\hat p_3)
	 \notag \\
	 	&\times j_\ell(qx)j_{\ell_2}(p_2 x)j_{\ell_3}(p_3 x)
	\notag 	\\
		=&4\pi \int \frac{p_2^2dp_2}{2\pi^2}\int \frac{p_3^2dp_3}{2\pi^2} \int x^2 dx  
		\notag \\
		&\times  j_0(qx)  j_{0}(p_2 x)j_{0}(p_3 x)  
	\notag 			\\
		=&4\pi \int \frac{p_2^2dp_2}{2\pi^2}\int \frac{p_3^2dp_3}{2\pi^2}  
		\frac{\pi}{8 p_2 p_3 q} \left(\frac{1}{\text{sgn}(p_2-p_3+q)}\right.
		\notag \\
		&\left.+\frac{1}{\text{sgn}(-p_2+p_3+q)}+\frac{1}{\text{sgn}(p_2+p_3-q)}-1\right),\label{A6}	
\end{align}
where we used the partial wave expansion for the plain waves in the third line.
The bracket in the last line reduces to a top hat filter for $|q-p_3|\leq p_3\leq q+p_3$.

\bibliography{sample.bib}{}
\bibliographystyle{unsrturl}

\end{document}